\documentclass[aps,twocolumn,
amsmath,prd,superscriptaddress,notitlepage,10pt
]{revtex4-1}

\usepackage[utf8]{inputenc}
\usepackage[T1]{fontenc}
\usepackage{dcolumn}
\usepackage{bm}
\usepackage{amsmath}
\usepackage{graphicx}
\usepackage{tabularx}
\usepackage{xfrac}
\usepackage{multirow}
\usepackage{rotating} 
\usepackage{makecell}
\usepackage{gensymb}
\usepackage{amssymb}
\usepackage{textgreek}
\usepackage[normalem]{ulem}
\usepackage[usenames, dvipsnames]{xcolor}
\usepackage{hyperref}
\hypersetup{colorlinks=true, linkcolor=blue, citecolor=blue, urlcolor=blue}
\usepackage{tabularx}
\usepackage{booktabs}

\begin{document}
\title{Synthesis of Mg$_2$IrH$_5$: A potential pathway to high-$T_c$ hydride superconductivity at ambient pressure}

%%%%%%%%%%%%%%%%%%%%%% AUTHOR LIST %%%%%%%%%%%%%%%%%%%%%%%%%%

\author{Mads F. Hansen}
\email{mhansen@carnegiescience.edu}
\affiliation{Earth and Planets Laboratory, Carnegie Institution for Science, 5241 Broad Branch Road NW, Washington, DC 20015, USA}

\author{Lewis J. Conway}
\affiliation{Department of Materials Science and Metallurgy, University of Cambridge, 27 Charles Babbage Road, Cambridge, CB3 0FS, UK}
\affiliation{Advanced Institute for Materials Research, Tohoku University, Sendai, 980-8577, Japan}

\author{Kapildeb Dolui}
\affiliation{Department of Materials Science and Metallurgy, University of Cambridge, 27 Charles Babbage Road, Cambridge, CB3 0FS, UK}

\author{Christoph Heil}
\affiliation{Institute of Theoretical and Computational Physics, Graz University of Technology, NAWI Graz, 8010 Graz, Austria}

\author{Chris J. Pickard}
\affiliation{Department of Materials Science and Metallurgy, University of Cambridge, 27 Charles Babbage Road, Cambridge, CB3 0FS, UK}
\affiliation{Advanced Institute for Materials Research, Tohoku University, Sendai, 980-8577, Japan}

\author{Anna Pakhomova}
\affiliation{European Synchrotron Radiation Facility, B.P.220, F-38043 Grenoble Cedex, France}

\author{Mohammed Mezouar}
\affiliation{European Synchrotron Radiation Facility, B.P.220, F-38043 Grenoble Cedex, France}

\author{Martin Kunz}
\affiliation{Advanced Light Source, Lawrence Berkeley National Laboratory, Berkeley, CA, USA}

\author{Rohit P. Prasankumar}
\affiliation{Intellectual Ventures, Bellevue, Washington, United States}

\author{Timothy A. Strobel}
\email{tstrobel@carnegiescience.edu}
\affiliation{Earth and Planets Laboratory, Carnegie Institution for Science, 5241 Broad Branch Road NW, Washington, DC 20015, USA}

%%%%%%%%%%%%%%%%%%%%%%%%%%%%%%%%%%%%%%%%%%%%%%%%%%%%%%%%%%%%%

\date{\today}

\begin{abstract}
Following long-standing predictions associated with hydrogen, high-temperature superconductivity has recently been observed in several hydride-based materials. Nevertheless, these high-$T_c$ phases only exist at extremely high pressures, and achieving high transition temperatures at ambient pressure remains a major challenge. Recent predictions of the complex hydride Mg$_{2}$IrH$_{6}$ may help overcome this challenge with calculations of high-$T_c$ superconductivity (65 K$~<~T_c~<~$ 170 K) in a material that is stable at atmospheric pressure. In this work, the synthesis of Mg$_{2}$IrH$_{6}$ was targeted over a broad range of $P$--$T$ conditions, and the resulting products were characterized using X-ray diffraction (XRD) and vibrational spectroscopy, in concert with first-principles calculations. The results indicate that the charge-balanced complex hydride Mg$_{2}$IrH$_{5}$ is more stable over all conditions tested up to \textit{ca}. 28 GPa. The resulting hydride is isostructural with the predicted superconducting Mg$_{2}$IrH$_{6}$ phase except for a single hydrogen vacancy, which shows a favorable replacement barrier upon insertion of hydrogen into the lattice. Bulk Mg$_{2}$IrH$_{5}$ is readily accessible at mild $P$--$T$ conditions and may thus represent a convenient platform to access superconducting Mg$_{2}$IrH$_{6}$ via non-equilibrium processing methods.% Finally, the critical factors influencing the calculated range of superconducting transition temperatures for this material are discussed.
\end{abstract}

\maketitle
Metallic hydrogen, and subsequently metallic hydrides, have been predicted to exhibit superconducting transition temperatures approaching room temperature.\cite{1968_Ashcroft,2004_Ashcroft} Recent computational and experimental studies indicate high-$T_{c}$ superconductivity (up to $\sim$250 K) in several hydrides including SH$_{3}$, LaH$_{10}$ and CeH$_9$ \cite{2015_Drozdov,Zulu_PhysRevLett.122.027001, Drozdov2019, 2018_Geballe,2021_Sun, Osmond_PhysRevB.105.L220502, Bhattacharyya2024, Chen_PhysRevLett.127.117001}. However, the highest-$T_{c}$ hydrides have only been observed at very high pressures (e.g., $>$100 GPa) and they are not recoverable to ambient pressure, rendering them non-viable for practical applications. While strategies such as doping and covalent stabilization have been proposed to reduce pressure requirements,\cite{Lilia_2022, Lucrezi2022,2023_Cataldo, SongPhysRevLett.130.266001} achieving high-$T_{c}$ hydrides at low-pressure conditions remains a grand challenge for conventional superconductivity. 

Recently, the complex hydride Mg$_{2}$IrH$_{6}$ was predicted to exhibit high-$T_{c}$ superconductivity at ambient pressure by several independent groups.\cite{2023_sanna,2023_Kapildeb,zheng2024prediction} The range of estimated transition temperatures varies between 65--170 K depending on computational details, and the material is predicted to be thermodynamically stable on the convex hull, or slightly metastable relative to other compounds at ambient pressure.\cite{2023_sanna,2023_Kapildeb,zheng2024prediction} Comprehensive structure searching over broad compositional space suggests that Mg$_{2}$IrH$_{6}$ is $\sim$60 meV/atom above the convex hull at ambient pressure, but could be accessible via an insulating Mg$_{2}$IrH$_{7}$ intermediate phase, which is predicted to be thermodynamically stable above 15 GPa.\cite{2023_Kapildeb} 

\begin{figure*}
    \centering
    \includegraphics[width = \linewidth]{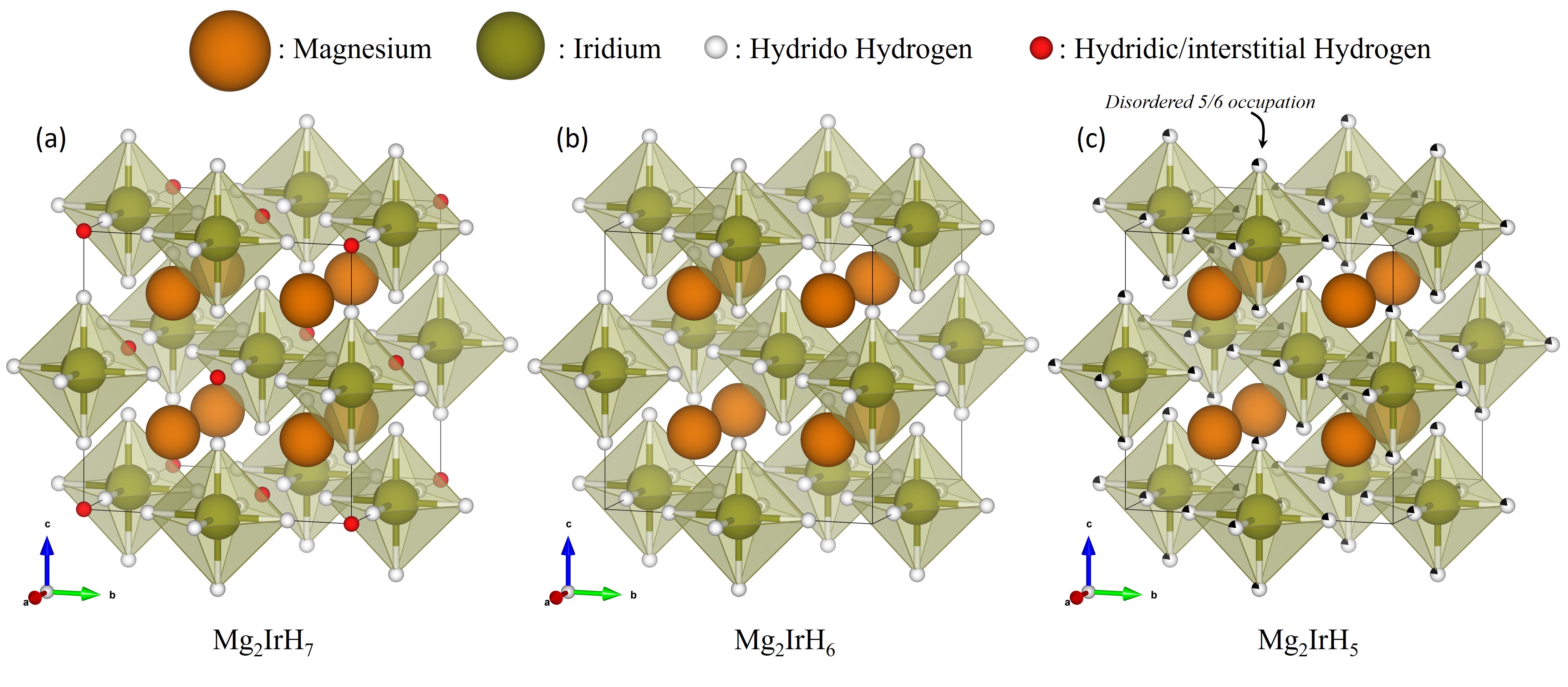}
    \caption{\textbf{a:} The crystal structure of Mg$_{2}$IrH$_{7}$. Red spheres indicate interstitial hydridic hydrogen which is more loosely bound than the hydrido complex hydrogen, represented by white spheres. \textbf{b:} The crystal structure of Mg$_{2}$IrH$_{6}$, \textbf{c:} The crystal structure of Mg$_{2}$IrH$_{5}$, which is isostructural with Mg$_{2}$IrH$_{6}$ but hydrogen octahedra are disordered with 5/6 site occupancy (black/white pie chart).}
    \label{fig:1}
\end{figure*}

Mg$_{2}$IrH$_{6}$ takes on the K$_{2}$PtCl$_{6}$ structure type (cubic $Fm\bar{3}m$) with octahedral [IrH$_{6}$]$^{3-}$ hydrido anions located at ($\sfrac{1}{2}$,$\sfrac{1}{2}$,$\sfrac{1}{2}$) and Mg$^{2+}$ cations at ($\sfrac{1}{4}$,$\sfrac{1}{4}$,$\sfrac{1}{4}$), as shown in Figure \ref{fig:1}. For the case of Mg$_{2}$IrH$_{7}$, which is predicted to be stable at high pressure, additional hydridic hydrogen occupies interstitial sites at (0,0,0). The synthesis of Mg$_{2}$IrH$_{6}$ was proposed by removing interstitial H from stable Mg$_{2}$IrH$_{7}$ based on its high mobility.\cite{2023_Kapildeb} 

High-$T_{c}$ superconductivity in Mg$_{2}$IrH$_{6}$ originates from hybridized H-1$s$, Mg-3$s$ and Ir-$e_{1g}$/$t_{2g}$ states with a high density of states at the Fermi level centered at a van Hove singularity, combined with strong electron--phonon coupling, predominantly from modes associated with hydrogen. 

In this Letter, we explore the synthesis of Mg$_{2}$IrH$_{6}$ over a broad range of $P$--$T$ conditions between $\sim$0--30 GPa and $\sim$450--2500 K where several phases are predicted to exist on or within $\sim$20 meV/atom of the convex hull. Experimental data over all tested conditions are consistent with the formation of insulating Mg$_{2}$IrH$_{5}$, which is face-centered cubic (FCC) in the same structure type as Mg$_{2}$IrH$_{6}$, but with disordered vacancies in the hydrido complex octahedra as shown in Figure \ref{fig:1}. The cubic Mg$_{2}$IrH$_5$ phase is likely stabilized by entropic contributions, not accounted for in previous calculations; however, Variable-Cell Nudged Elastic Band (VCNEB) calculations suggest that this phase may serve as a platform to access superconducting Mg$_{2}$IrH$_{6}$ via H insertion.

\begin{figure}
    \centering
    \includegraphics[width  = \linewidth]{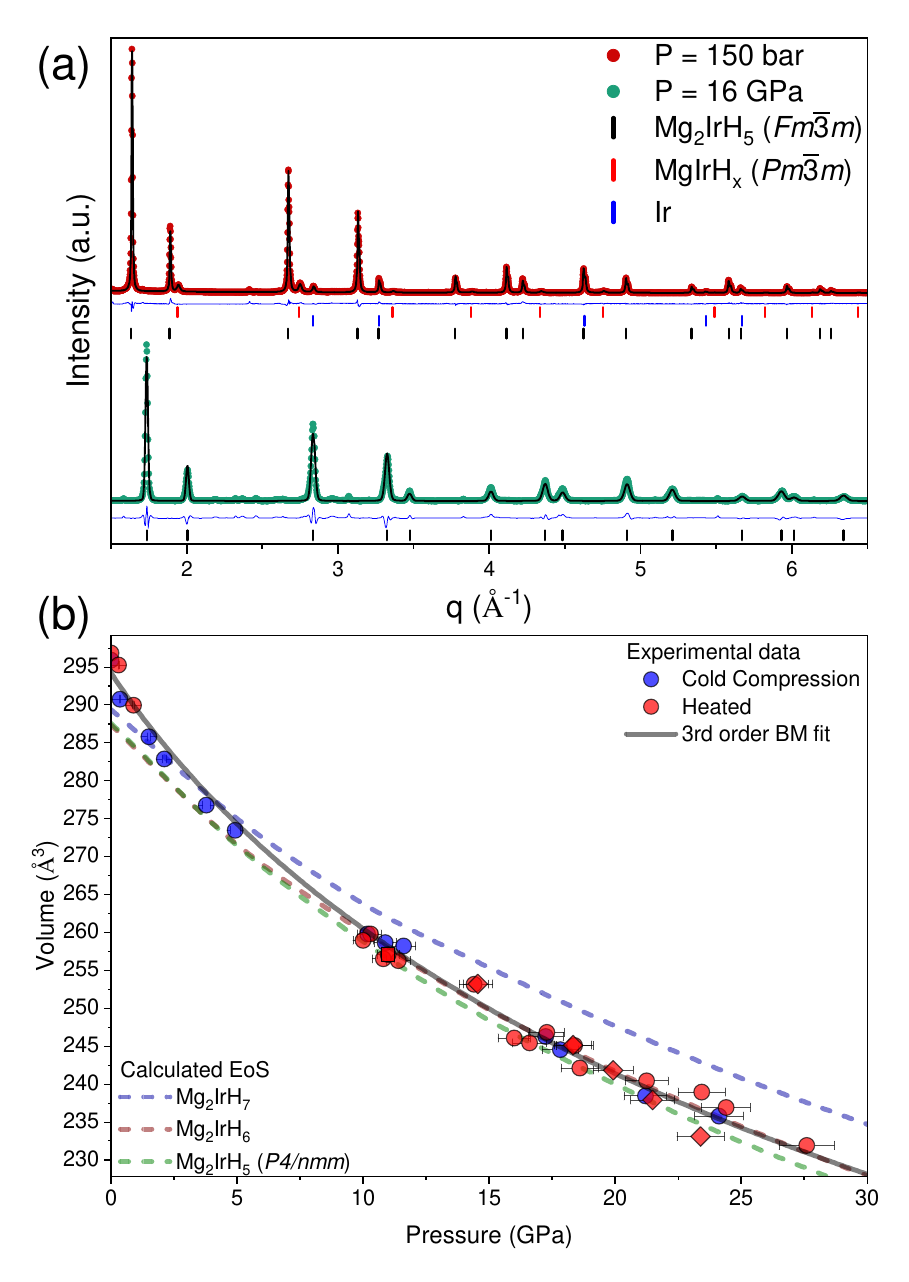}
    \caption{\textbf{a:} Powder XRD data (points) with Rietveld refinements (black lines) and residuals (blue lines) for samples synthesized at 150 bar (measured at 1 bar) in an autoclave, and 16 GPa in a DAC (measured at 16 GPa) after heating. Tick marks indicate different phases. Trace impurity phases including Mg$_{6}$Ir$_{2}$H$_{11}$ are present in the patterns. \textbf{b:} Pressure--volume relation for the FCC compound. Blue circles represent cold-compression (no heating) of the Mg$_2$IrH$_5$ precursor synthesized at low pressure, while red symbols indicate the FCC unit cell volume after heating at various conditions and during decompression (see the SM for probed conditions). Circles are autoclave-type precursors, rhombi represent the Mg$_{3}$Ir intermetallic precursor, and squares indicate elemental precursors. The data are fitted with a 3rd-order Birch-Murnaghan equation of state (EOS). Calculated DFT equations of state are shown as dashed lines.}
    \label{fig:2}
\end{figure}

Following the predicted synthetic pathway to Mg$_{2}$IrH$_{6}$ by forming Mg$_{2}$IrH$_{7}$ above 15 GPa, high-pressure diamond anvil cell (DAC) experiments were conducted using an elemental 2Mg + Ir mixture with high-density H$_{2}$ fluid as a reagent and pressure-transmitting medium, as well as experiments using intermetallic Mg$_3$Ir or Mg--Ir hydride precursors, as detailed in the Supplemental Material (SM). Initial experiments were conducted at 10--15 GPa using infrared laser heating with \textit{in situ} synchrotron X-ray diffraction (Table S1). Upon heating the elemental mixture to moderate temperatures near $\sim$800 K, samples rapidly convert (nearly phase pure) to a new FCC lattice with $a \approx$ 6.37~\AA~ at 10 GPa, in close agreement with DFT predictions for Mg$_{2}$IrH$_{6/7}$ ($a \approx$ 6.40~\AA~  at the same pressure). Given that hydrogen has the lowest scattering cross section of all elements, it is difficult to detect by X-ray scattering. Rietveld refinement of powder data at various pressures (Figure \ref{fig:2}), as well as energy-dispersive X-ray spectroscopy (EDS) measurements on recovered samples (see SM), indicate the formation of the FCC Mg$_{2}$Ir sublattice found in both Mg$_2$IrH$_{6/7}$ phases. Nevertheless, establishing the precise hydrogen content based on comparison with calculated unit cells is problematic given the small volume differences between the compounds and the uncertain variation between experiment and the static DFT calculations.

%Distinguish calc and exp, change colors for calculations,

To further probe the influence of synthetic conditions on unit cell volume, additional experiments were conducted up to 28 GPa using temperatures ranging from $\sim$800--2500 K. For all conditions tested, the majority product was an FCC phase with lattice parameters that closely track calculations for Mg$_2$IrH$_{6}$ as a function of pressure, with slightly larger deviations from Mg$_2$IrH$_{7}$ (Figure \ref{fig:2}b and SM). In fact, it was found that an FCC phase matching the same volume trend could be produced from the elements at low hydrogen pressures near 150 bar using an autoclave-type reactor system (see SM). A similar reaction product was reported in ref. \cite{1995_Bonhomme_thesis} with the preliminary interpretation of FCC Mg$_{2}$IrH$_{5}$, although characterization was limited to XRD probing the fluorite-type metal sublattice, and IR spectroscopy showing a strong absorbance at 1852 cm$^{-1}$. An FCC M$_{2}$IrH$_{5}$ phase containing [IrH$_5$]$^{4-}$ units is well established for M = (Ca, Sr, Eu) using neutron diffraction.\cite{2012_Barsan,1980_Moyer,1981_Zhuang,2003_Kohlmann,1971_Moyer,2009_Kohlmann} The calculated $P$--$V$ trends for different polymorphs of Mg$_{2}$IrH$_{5}$ also closely match the experimental volume data (Figure \ref{fig:2}b). In every case examined, the resulting unit cell volume follows a continuous trend between 0--28 GPa across various heating conditions, with no clear indications of a deviation in hydrogen stoichiometry. %(i.e., the same approximate composition is always formed).DFT calculation show that the unit cell volumes of H5/6/7 are very similar, with minor increases with increasing H content. The H content from experiment cannot unambiguously be determined from volume, but all 
The calculated volumes for Mg$_{2}$IrH$_{x}$ ($x$ = 5--7) are all potentially consistent with experimental diffraction results (Mg$_{2}$IrH$_{7}$ shows the largest deviation of $\sim$3 \%), but the precise composition cannot be unambiguously determined using this approach.

To probe the hydrogen sublattice and reveal information about the local H environments of the reaction product(s), Raman and infrared spectroscopic measurements were performed on the FCC samples after synthesis at various $P$--$T$ conditions. Subtle differences in local Ir--H bonding environments will be reflected in the modes and frequencies present in vibrational spectra, which can be used to help establish metal coordination and possible variations in hydrogen content.

\begin{figure*}[ht!]
    \centering
    \includegraphics[width = \linewidth]{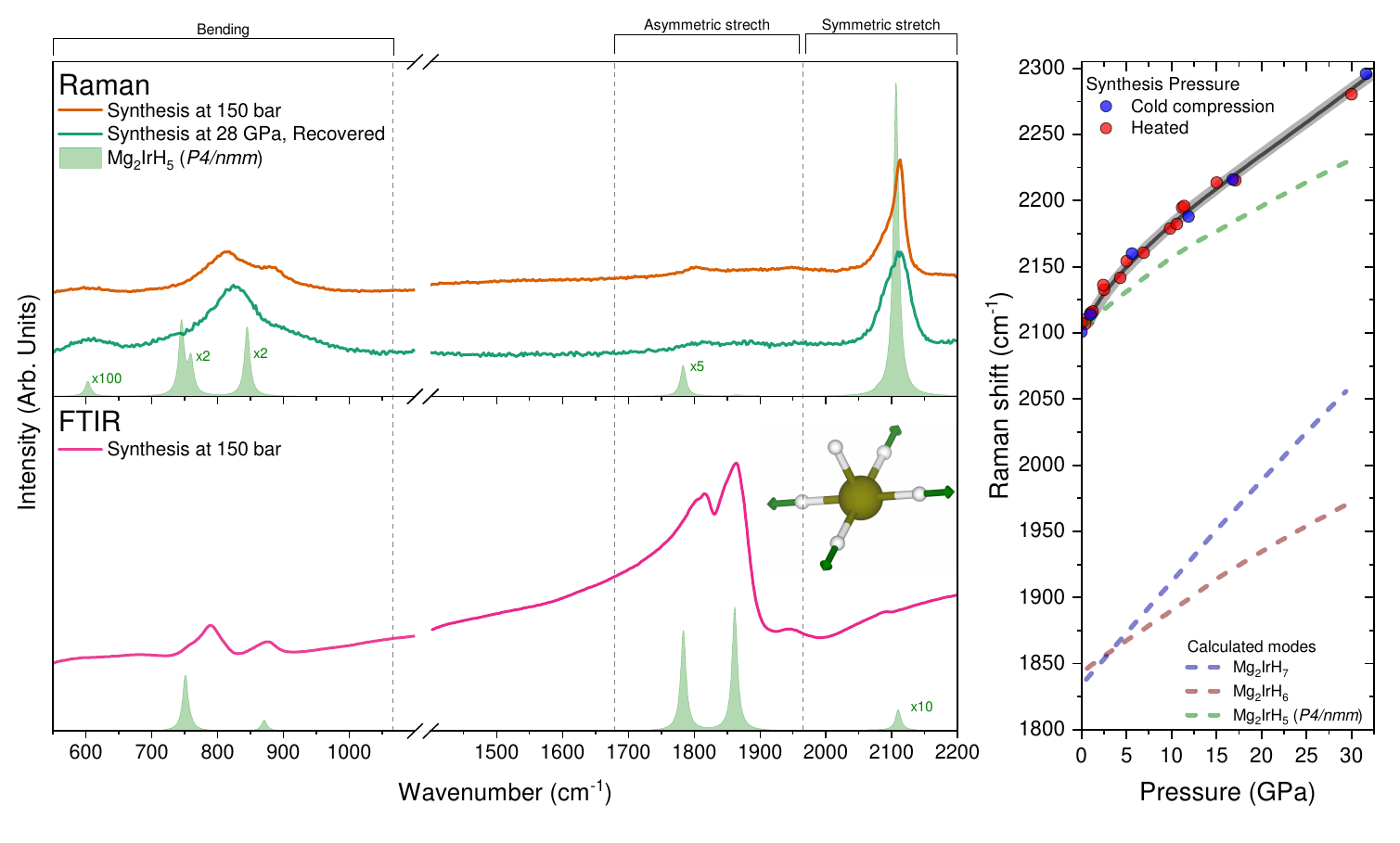}
    \caption{Spectroscopic data for FCC samples produced at different $P$--$T$ conditions. %Mg$_{2}$IrH$_{5}$. 
    \textbf{a:} Raman spectra of FCC samples taken at ambient pressure for samples synthesized at 150 bar and 28 GPa, as well as the FTIR spectrum of a sample synthesized at 150 bar. All spectra shown are at ambient pressure. The region from 1200--1500 cm$^{-1}$ is removed due to Raman signal from the diamonds (see SM for full spectra). The experimental data are only consistent with calculations for Mg$_{2}$IrH$_{5}$ (green peaks). The inset displays the symmetric stretching mode of the [IrH$_{5}$]$^{4-}$ unit. \textbf{b:} Pressure dependence of the symmetric stretching mode of [IrH$_{5}$]$^{4-}$ for samples synthesized at different conditions. Pressure-dependant calculations of different phases are shown for comparison.
    }
    \label{fig:3}
\end{figure*}

Figure \ref{fig:3}a compares example spectra obtained from a sample produced using autoclave synthesis at 150 bar with a recovered DAC sample synthesized at 28 GPa and $\sim$1700 K. The spectra obtained from these two samples, which were produced across the range of tested synthetic conditions, show virtually identical Raman features including a strong Ir--H stretching mode near $\sim$2100 cm$^{-1}$, Ir--H bending modes near $\sim$800 cm$^{-1}$, as well as lattice phonons at lower energy (see full spectra in the SM). Qualitatively, the spectra show similar stretching and bending modes observed in previous reports of other complex hydrides that contain MH$_5$ or MH$_6$ complexes.\cite{1997_Parker, Parker1998_A803802C, PARKER2010215, Barsan2012} These same spectral features were observed for all samples synthesized over the entire range of $P$--$T$ conditions studied (see Table S1 in SM), suggesting that the local hydrogen environment remains constant within a single structure with continuous volume compressibility based on XRD. In particular, the pressure dependence of the Ir--H stretching mode observed at $\sim$2100 cm$^{-1}$ is shown in Figure \ref{fig:3}b. This mode follows the same trend line for all samples synthesized over the entire range of $P$--$T$ conditions studied.

To gain more insight into the nature of hydrogen within the FCC hydride phase, the Raman spectra of possible Mg$_2$IrH$_x$ structures were calculated for comparison with experimental data. Ir--H stretching/bending frequencies are sensitive to the local coordination environment, providing a key diagnostic to distinguish different molecular complexes e.g., [IrH$_5$]$^{4-}$ vs. [IrH$_6$]$^{3-}$.
%$C_{4v}$ $O_{h}$
Cubic Mg$_2$IrH$_6$ has four Raman-active modes with $A_{1g}$ calculated at $\sim$1850 cm$^{-1}$, $E_g$ at $\sim$1560 cm$^{-1}$, and two $T_{2g}$ modes at $\sim$750 and $\sim$220 cm$^{-1}$ at 0 GPa. Mg$_2$IrH$_7$ exhibits the same Raman modes with small perturbations in frequency (i.e., shifts by $\leq$10 cm$^{-1}$), with the exception of the $T_{2g}$ Ir--H bending mode at $\sim$950 cm$^{-1}$, which is stiffened by roughly 200 cm$^{-1}$ due to the incorporation of interstitial hydrogen on the octahedral sites. The experimentally observed Raman frequencies generally show poor agreement with calculations for Mg$_2$IrH$_6$ and Mg$_2$IrH$_7$ (See SM). In particular, the intense peak observed at 2108 cm$^{-1}$, which is assigned to Ir--H symmetric stretching based on Raman activity and comparison with other hydrido complexes,\cite{1997_Parker, Parker1998_A803802C, PARKER2010215, Barsan2012} is more than 250 cm$^{-1}$ stiffer than the symmetric A$_{1g}$ stretching modes calculated for Mg$_2$IrH$_6$ and Mg$_2$IrH$_7$. %\TS{(Benchmark calculations with the same approach for Ca$_2$IrH$_5$ show calculated stretching frequencies $\sim$XX cm$^{-1}$ from experimental values???)}

%Distinguish calc and exp, change colors for calculations, Add x5 for asym stretch raman
Calculating the vibrational spectra of FCC Mg$_{2}$IrH$_{5}$ is not feasible since the [IrH$_5$]$^{4-}$ units, which are disordered, are not compatible with the symmetry of an ordered FCC model. Therefore, we calculated the Raman spectra of several ordered approximations of cubic Mg$_{2}$IrH$_{5}$ with lower symmetry. With this approach, the molecular modes associated with Ir--H stretching and bending are captured well and are largely independent of the specific lattice (Figure \ref{fig:3}), whereas the lattice phonons specific to the FCC structure are not expected to be portrayed as accurately. The experimentally observed high-frequency Ir--H stretching mode is described well by calculated spectra of several ordered models for Mg$_{2}$IrH$_{5}$ (see SM). This high-frequency vibration is not described by the octahedral units found in Mg$_{2}$IrH$_{6/7}$, suggesting that the experimental FCC compound contains [IrH$_{5}$]$^{4-}$ complexes. This assignment is strengthened by overall agreement with calculated Ir--H bending frequencies, and while the intensity of the low-frequency lattice phonons are not captured well by the calculations, these frequencies show qualitative agreement (see SM). The [IrH$_5$]$^{4-}$ complex is further confirmed by the infrared absorption spectrum (the FCC phase transmits IR light and displays semiconductor electrical transport behavior), which shows asymmetric Ir--H stretching at $\sim$1800 cm$^{-1}$ and bending modes around $\sim$800 cm$^{-1}$ (Figure \ref{fig:3}a), the former being consistent with the previous report.\cite{1995_Bonhomme_thesis} In addition, the observed Raman/IR spectra agree well with the analogous FCC Ca$_2$IrH$_5$ and Sr$_2$IrH$_5$ compounds, as well as other known hydrido complexes containing [MH$_{5}$] motifs.\cite{1981_Zhuang, Parker1998_A803802C, PARKER2010215,2006_Gilson, Barsan2012}

Thus, the pressure dependence of the symmetric Ir--H stretching mode in Figure \ref{fig:3}b for samples synthesized over a range of $P$--$T$ conditions, following a singular trend for pressures up to 30 GPa, indicates that [IrH$_{5}$]$^{4-}$ complexes persist over all measured conditions. Based on XRD observations showing a single $P$--$V$ trend and Raman/IR observations that show a single chemical environment, Mg$_{2}$IrH$_{5}$ appears to be the only FCC compound produced over all conditions tested. This composition is further supported by thermogravimetric analyses performed on spectroscopically equivalent samples produced at low pressure (see SM).

As mentioned above, computational models of Mg$_{2}$IrH$_{5}$ exhibit lower symmetry than cubic due to fact that the IrH$_{5}$ units cannot occupy an ordered FCC lattice, which contains six equivalent H sites. In experiment, however, substitutional disorder allows for a cubic unit cell (time--space average observation), where a hydrogen vacancy is distributed equally among the IrH$_{6}$ octahedra that would be present in Mg$_{2}$IrH$_{6}$. Mg$_{2}$IrH$_{5}$ can be described as the same structure with a hydrogen occupancy of $\sfrac{5}{6}$, as indicated in Figure \ref{fig:1}. For other isostructural M$_{2}$IrH$_{5}$ (M = Ca, Sr, Eu) compounds, ordering of the IrH$_{5}$ complex occurs at lower temperature, and the disordered cubic hydrides transform to ordered tetragonal structures, analogous to the current calculations.\cite{1980_Moyer,1981_Zhuang} 

%The synthesized compound has a reddish-brown/black appearance in reflected visible light and transmits infrared light and is clearly not metallic as predicted for Mg$_{2}$IrH$_{6}$, which further supports the formation of charge-balanced Mg$_{2}$IrH$_{5}$.
%\TS{Experimental absorbance measurements indicate an optical band gap near near XX eV, which compares favorably with DFT predictions of XX eV. The electrical resistivity, measured on a 95\% phase pure pressed powder pellet, was determined to be XX $\ohm \cdot m$, and shows semiconducting temperature dependence.}   

\begin{figure*}[ht!]
    \centering
    \includegraphics[width = \linewidth]{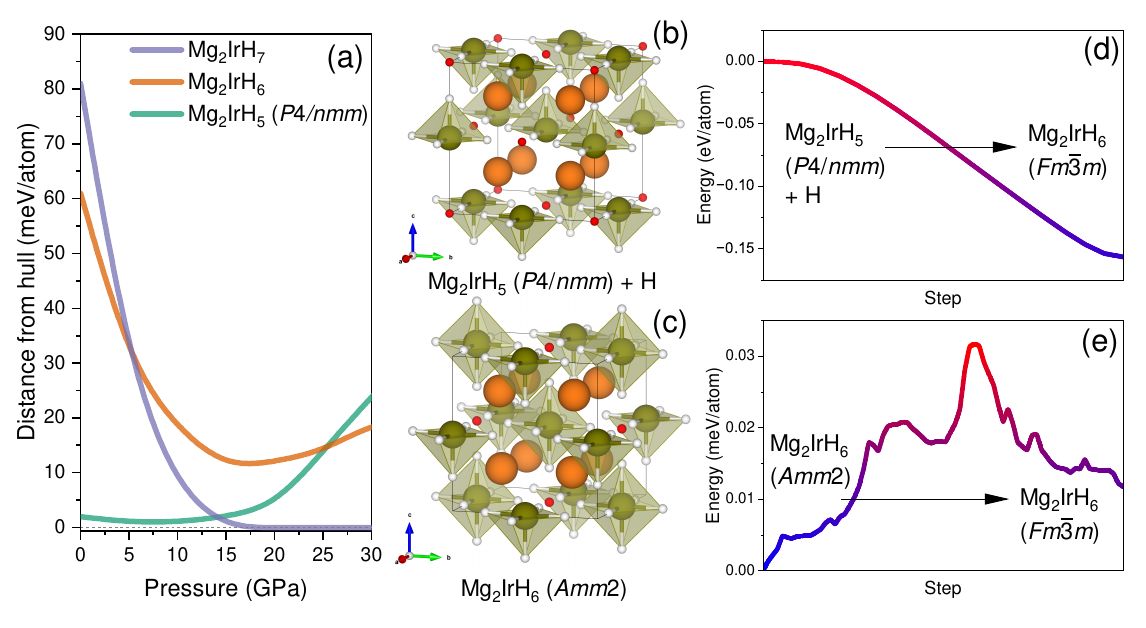}
    \caption{\textbf{a:} Calculated convex hull distances for FCC Mg$_{2}$IrH$_{6/7}$ and Mg$_{2}$IrH$_{5}$ ($P$4/$nmm$) as a function of pressure. The structures of Mg$_{2}$IrH$_{5}$ ($P$4/$nmm$) + interstitial H displayed in the same cubic basis as the structure in figure \ref{fig:1} to ease comparison. (\textbf{b}) and Mg$_{2}$IrH$_{6}$ ($Amm$2) (\textbf{c}), both used as a starting points for VCNEB calculations. \textbf{d:} VCNEB trajectory starting from Mg$_{2}$IrH$_{6}$ ($P$4/$nmm$) + H to Mg$_{2}$IrH$_{6}$ ($Fm\bar{3}m$) \textbf{e:} VCNEB trajectory starting from Mg$_{2}$IrH$_{6}$ ($Amm$2) to Mg$_{2}$IrH$_{6}$ ($Fm\bar{3}m$).}
    \label{fig:4}
\end{figure*}

Experimental data indicate the formation of Mg$_{2}$IrH$_{5}$ for synthesis conditions between $\sim$0--30 GPa, whereas previous DFT calculations suggested that Mg$_{2}$IrH$_{7}$ should be stable above 15 GPa. This discrepancy can most likely be attributed to effects associated with finite temperature where the static approximation is not sufficient to describe the empirical phase behavior of Mg$_{2}$IrH$_{5}$. Configurational entropy associated with disordered IrH$_5$ units, which was not previously accounted for, likely plays a significant role in stabilizing this phase over a broad range of $P$--$T$ conditions. According to previous calculations, Mg$_{2}$IrH$_{7}$ is located on the convex hull above 15 GPa (Figure \ref{fig:4}a) where Mg$_{2}$IrH$_{6}$ exhibits a minimum hull distance of $\sim$10 meV/atom.\cite{2023_Kapildeb} Different forms of ordered Mg$_{2}$IrH$_{5}$ are only slightly above the hull at 0 GPa and become destabilized at pressures above \textit{ca.} 15 GPa. Nevertheless, the consistent experimental observation of Mg$_{2}$IrH$_{5}$ suggests that when entropic contributions are taken into account at finite temperature, including the disordered nature of Mg$_{2}$IrH$_{5}$ ($S_{\rm config}=k_{\rm B} \ln(6)$ per F.U., which gives 5.8 meV/atom at 300 K), this phase likely becomes the thermodynamic ground state for pressures up to at least $\sim$30 GPa.

While cubic Mg$_{2}$IrH$_{5}$ is an insulating phase (DFT band gap in $P$4/$nmm$ is $\sim$0.65 eV), its structure is exceptionally close to the predicted high-$T_c$ superconductor Mg$_{2}$IrH$_{6}$. The structures possess identical metallic sublattices and only a single hydrogen needs to be added per formula unit to obtain the superconducting phase. Mg$_{2}$IrH$_{6}$ is dynamically stable at ambient conditions with a moderate energy distance above the convex hull, and might be accessible via hydrogen addition to the synthesized Mg$_{2}$IrH$_{5}$ phase. To investigate this possibility, we calculated kinetic barriers for the formation of Mg$_{2}$IrH$_{6}$ upon hydrogen insertion using VCNEB calculations,\cite{VCNEB,USPEX} as shown in Figure \ref{fig:4}d-e. Starting from Mg$_{2}$IrH$_{5}$ ($P$4/$nmm$) with additional interstitial hydrogen placed in the lattice at the 2$c$ Wyckoff position, our calculations show that the system smoothly transitions to the FCC Mg$_{2}$IrH$_{6}$ structure with no indication of an energetic barrier (Figure \ref{fig:4}d). Thus, metastable implantation of interstitial H within Mg$_{2}$IrH$_{5}$ appears to be a favorable route to obtain the target compound. Nevertheless, Dolui et al. predicted another $Amm$2 polymorph of Mg$_{2}$IrH$_{6}$ that is slightly more stable than the $Fm\overline{3}m$ phase at low pressure and could also potentially form during hydrogenation of Mg$_{2}$IrH$_{5}$.\cite{2023_Kapildeb} For the case of $Fm\overline{3}m$ Mg$_{2}$IrH$_{6}$ formation from the $Amm$2 phase, VCNEB calculations (Figure \ref{fig:4}e) suggest a kinetic barrier of 32 meV/atom ($\sim$370 K), consistent with quasiharmonic free-energy calculations that show favorability of the FCC phase at high temperature.\cite{2023_Kapildeb} Based on these calculations, Mg$_{2}$IrH$_{5}$, which can be scaled in bulk at low-pressure conditions, represents a promising platform for accessing Mg$_{2}$IrH$_{6}$ using non-equilibrium methods such as deposition or implantation.

Finally, it is important to note that the target phase Mg$_{2}$IrH$_{6}$ has a wide range of predicted $T_{c}$s.\cite{2023_sanna,2023_Kapildeb,zheng2024prediction} This is largely due to the fact that there is a particular band at the Fermi level with high DOS and a van Hove singularity, making calculations very sensitive to the chosen smearing and reciprocal grid parameters. In the case of highly converged calculations with very dense grids, a robust $T_c$ range between $\sim$90--120 K is reproduced independently across multiple methods with the main differences originating from (i) employing the isotropic approximation or working fully anisotropically, (ii) whether or not scattering processes away from the Fermi level are taken into account,\cite{2024_Lucrezi} and (iii) how the screened Coulomb interaction is incorporated.\cite{2023_Pellegrini} Thus, Mg$_{2}$IrH$_{6}$ remains a strong candidate for high-$T_c$, low-pressure hydride superconductivity, and the nearly isostructural Mg$_{2}$IrH$_{5}$ compound can serve as an important starting point to access this remarkable phase.

\section*{Acknowledgement}
The authors would like to thank Emma Bullock for assistance with SEM/EDX measurements.
This work was supported by the Deep Science Fund of Intellectual Ventures. We acknowledge the European Synchrotron Radiation Facility (ESRF) for provision of synchrotron radiation facilities at beamline ID27 (CH-6822).\cite{2023_CH-6822_ESRF1} This research used resources of the Advanced Light Source (beamline 12.2.2), which is a DOE Office of Science User Facility under contract no. DE-AC02-05CH11231. Portions of this work were performed at HPCAT (Sector 16), Advanced Photon Source (APS), Argonne National Laboratory. HPCAT operations are supported by DOE-NNSA’s Office of Experimental Sciences. The Advanced Photon Source is a U.S. Department of Energy (DOE) Office of Science User Facility operated for the DOE Office of Science by Argonne National Laboratory under Contract No. DE-AC02-06CH11357.

\clearpage 

\onecolumngrid
\begin{center}
\textbf{Supplemental Material for ``Supplemental Material - Synthesis of Mg$_{2}$IrH$_{5}$: A potential pathway to high-$T_{c}$ hydride superconductivity at ambient pressure"}
\end{center}

%\documentclass[aps,onecolumn,
%amsmath,prd,superscriptaddress,notitlepage,10pt
%]{revtex4-1}

%\usepackage[utf8]{inputenc}
%\usepackage[T1]{fontenc}
%\usepackage{dcolumn}
%\usepackage{bm}
%\usepackage{amsmath}
%\usepackage{graphicx}
%\usepackage{tabularx}
%\usepackage{xfrac}
%\usepackage{multirow}
%\usepackage{rotating} 
%\usepackage{makecell}
%\usepackage{gensymb}
%\usepackage{amssymb}
%\usepackage{textgreek}
%\usepackage[normalem]{ulem}
%\usepackage[usenames, dvipsnames]{xcolor}
%\usepackage{hyperref}
%\hypersetup{colorlinks=true, linkcolor=blue, citecolor=blue, urlcolor=blue}
%\usepackage{tabularx}
%\usepackage{booktabs}

%Color commands to be used for commenting
%\newcommand\blue[1]{{\color{blue}#1}}
%\newcommand\TS[1]{{\color{BurntOrange}#1}}
%\newcommand\org[1]{{\color{orange}#1}}
%\newcommand\MF[1]{{\color{blue}#1}}
%\newcommand\LJC[1]{{\color{purple}#1}} %For Lewis

\renewcommand{\thefigure}{S\arabic{figure}}
\renewcommand{\thetable}{S\Roman{table}}
\setcounter{figure}{0}

\section*{Methods}
\paragraph*{Synthesis:}
High-pressure experiments were conducted using symmetric and BX90 diamond anvil cells (DAC) with 300-600 $\mu$m culet anvils and Re gaskets indented to $\sim$40 $\mu$m and laser drilled for the sample chamber. Mg--Ir precursors were loaded within an inert Ar glovebox and then the sample chamber was filled with high-density H$_2$ fluid at $\sim$1.5 kbar, which served as a reagent and pressure-transmitting medium. %Pressure was determined using the ruby fluorescence and diamond Raman scales \TS{(REFS)} for optical measurements, and the Ir equation of state \TS{(REFS)} for \textit{in situ} X-ray diffraction measurements.

Three different starting precursor materials were used for high-pressure experiments: 1) well-mixed 2Mg+Ir elemental powders (pressed as thin pellets); 2) a Mg$_3$Ir intermetallic phase produced by heating the elements at 900 K under Ar; 3) FCC Mg$_{2}$IrH$_{5}$ powder produced in a low-pressure autoclave-type assembly. These precursor samples were heated in molecular hydrogen over a broad range of P--T conditions.

To synthesize Mg$_{2}$IrH$_{5}$ outside the DAC, we mixed Mg (Sigma Aldrich 99.5\%) and Ir (Sigma Aldrich 99.9\%) in a 2:1 molar ratio and heated it at 450\textdegree{}C for two weeks under 100--250 bars of hydrogen (UHP Airgas) within a stainless steel autoclave-type reactor. The resulting product is FCC Mg$_{2}$IrH$_{5}$ (typically 80--90 wt\%), with additional impurities including a  $Pm\overline{3}m$ MgIrH$_x$ alloy, unreacted FCC Ir, and monoclinic Mg$_6$Ir$_3$H$_{11}$.\cite{2002_Cerny} Trace MgH$_2$ was also observed in some experiments of shorter duration.

Samples were compressed to various target pressures between 0--30 GPa, and one- or two-sided laser heating was performed using laser heating systems (typically 100 W, 1064 nm YAG laser) and temperature was estimated by thermal emission. In most cases, a chemical reaction was observed at low laser powers at the threshold of coupling ($\sim$800 K). Given the motion of fluid hydrogen during the heating process, laser coupling, and thus temperature, fluctuated with time. Instantaneous temperature measurements at low and high laser power ranged between $\sim$800--2500 K. The various experimental schemes are summarized in Table \ref{SI:table_1}.
\\

\begin{figure*}[ht!]
    \centering
    \includegraphics[width = .8 \linewidth]{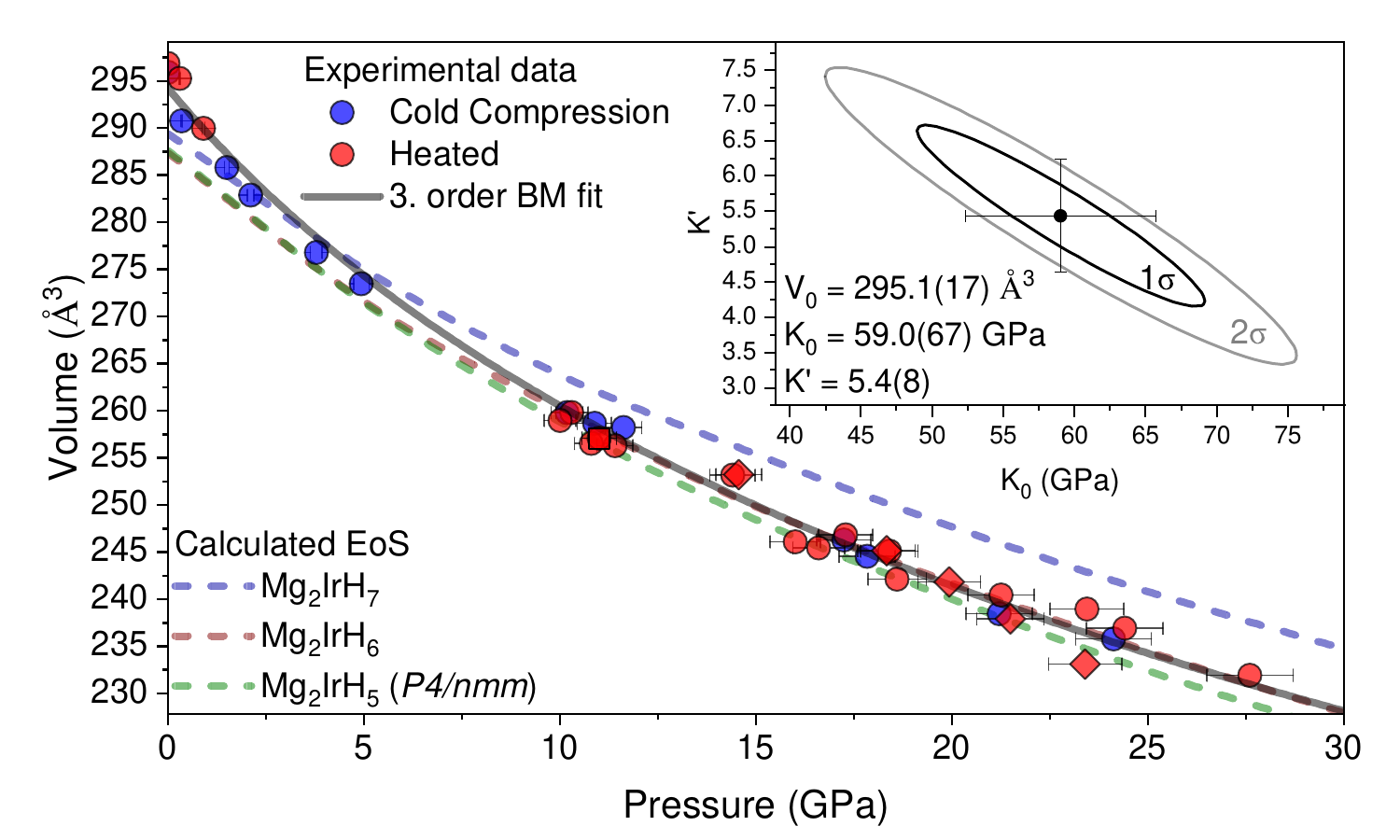}
    \caption{Pressure--volume relation for the FCC compound. Blue circles represent cold-compression (no heating) of the Mg$_2$IrH$_5$ precursor synthesized at low pressure, while red symbols indicate the FCC unit cell volume after heating at various conditions and during decompression. Circles are autoclave-type precursors, rhombi represent Mg$_{3}$Ir intermetallic precursors, and squares indicate elemental precursors. The data are fitted with a 3rd-order Birch-Murnaghan equation of state (EOS). Calculated DFT equations of state are shown as dashed lines. The inset shows the covariance ellipsoid of K$_{0}$ and K' with uncertainty at one and two standard deviations.}
    \label{SI:fig:EOS}
\end{figure*}

\paragraph*{X-ray diffraction:}
%X-ray diffraction as carried out in symmetric and BX90 type DACs loaded with a precursor and hydrogen as pressure transmitting medium (PTM). Rhenium was used as gasket material with a preindent to $\sim$40 $\mu$m. The gas loading was done in a home build system with a loading pressure of $\sim$ 20 kpsi. Laser heating was performed at the ESRF, using a 100 W, 1064 nm YAG laser. 
Angular-dispersive, monochromatic X-ray diffraction measurements were carried out in transmission geometry at ID27 at the ESRF with a wavelength of 0.3738 \AA~using a Dectris Eiger 9M CdTe detector \cite{2023_CH-6822_ESRF}; at beamline 12.2.2 \cite{Kunz:xd5008} at the Advanced Light Source using a wavelength of 0.4959 \AA~using a Pilatus Si detector; and at 16IDB (HPCAT) at the Advanced Photon Source using a wavelength of 0.4066 \AA~using a Pilatus CdTe detector. The sample-to-detector distances and detector geometries were calibrated using CeO$_2$ and LaB$_6$ standards, in conjunction with the Dioptas software \cite{Dioptas}, which was also used for data reduction. For \textit{in situ} diffraction measurements, pressure determination was established using the equation of state (EOS) of Ir from small regions of unreacted sample near the laser heating spot. \cite{2019_Monteseguro,2019_Yusenko,2022_Han,Kumar_2016,2016_Burakovsky,2020_Moseley,2000_Cerenius} The EOS from Montesguro et al. was found to be most suitable for the pressure range examined here. \cite{2019_Monteseguro} Rietveld refinement was performed using the Jana2020 software \cite{Jana} for cases where data showed suitable powder averaging statistics. In cases with spotty patterns or texture, Le Bail refinement was used to establish lattice parameters. 
We also established the P--V equation of state for Mg$_{2}$IrH$_{5}$ by fitting a third-order Birch--Murnaghan (BM) equation using the EoSFit software. \cite{EOSfit} In Figure \ref{SI:fig:EOS} we show the equation of state fit from the main text with an inset showing the covariance ellipsoid and the obtained equation of state parameters.
Diffraction patterns from samples synthesized in the autoclave-type system were measured using either a Bruker D2 PHASER (λ = 1.5406 Å) equipped with a LYNXEYE 1D detector, or a Bruker D8 diffractometer (λ = 1.5406 Å) equipped with a 2mm collimator and VANTEC-500 area detector.

\begin{table}[h]
    \caption{Summary of conditions for DAC experiments carried out in the synthesis of Mg$_{2}$IrH$_{x}$}
    \label{SI:table_1}
    \begin{tabular}{  l  p{9.4cm} }
        \toprule
        \textbf{Cell \#} & \textbf{Experimental details} \vspace{0.1cm} \\ \hline
        Cell 1      & Mg$_{2}$IrH$_{5}$ synthesized in autoclave compressed in H$_2$ to $\sim$30 GPa with \textit{in situ} XRD. The diamonds failed during compression.\vspace{0.1cm} \\ \hline
        
        Cell 2      & Two separate sample pellets (autoclave Mg$_{2}$IrH$_{5}$ and intermetallic Mg$_3$Ir) compressed in H$_2$ up to $\sim$27 GPa with \textit{in situ} XRD. The sample was then laser heated at high pressure, then  decompressed with \textit{in situ} XRD. Raman spectra were collected after heating. The diamonds failed during decompression.\vspace{0.1cm} \\ \hline
        
        Cell 3      & Autoclave Mg$_{2}$IrH$_{5}$ compressed in H$_2$ up to $\sim$24 GPa with \textit{in situ} XRD. The sample was then laser heated at high pressure, then  decompressed with \textit{in situ} XRD and Raman.\vspace{0.1cm} \\ \hline
        
        Cell 4      & 2Mg:1Ir elemental pellet compressed in H$_2$ to $\sim$11 GPa, the laser heated with \textit{in situ} XRD. Raman spectra were collected on decompression. \vspace{0.1cm} \\ \hline
        
        Cell 5      & Autoclave Mg$_{2}$IrH$_{5}$ compressed in H$_2$ up to $\sim$16 GPa with \textit{in situ} XRD. The sample was then heated using different laser powers to examine the influence of temperature dependence on the reaction product. XRD and Raman spectroscopy was measured at the target pressure. \vspace{0.1cm} \\ \hline
       
        Cell 6      & Autoclave Mg$_{2}$IrH$_{5}$ compressed in H$_2$ up to $\sim$1.5 GPa with \textit{in situ} XRD and laser heating. The sample was then brought to 10.5 GPa and another portion of the sample was laser heated again. XRD was measured at both pressure points, Raman spectroscopy was measured at $\sim$10.5 GPa. \vspace{0.1cm} \\ \hline

        Cell 7      & Autoclave Mg$_{2}$IrH$_{5}$ compressed to $\sim$1 GPa in H$_2$, laser heated and measured Raman spectroscopy.\vspace{0.1cm} \\ \hline

        Cell 8      & Autoclave Mg$_{2}$IrH$_{5}$ compressed to $\sim$30 GPa in H$_2$, laser heated and measured Raman spectroscopy upon decompression.\vspace{0.1cm} \\ \hline
        \bottomrule
    \end{tabular}
\end{table}

\paragraph*{Raman spectroscopy:}
Unpolarized Raman spectra were measured using a 532 nm laser for excitation focused through a 20$\times$ long-working-distance objective lens at low power ($\sim$ 5 mW) and a Princeton instruments SP2750 spectrograph with a 750 mm focal length. The counting time was up to 120 s averaging 10 times to obtain a high signal to noise ratio. The pressure was determined using ruby fluorescence \cite{1986_Mao} or diamond Raman \cite{Raman_diamond}. Most samples were measured upon decompression after laser heating in the presence of molecular H$_2$. One sample synthesized by laser heating in DAC at 28 GPa and $\sim$1700 K was recovered and reloaded in a DAC, and reloaded with Ar to remove hydrogen Raman features. This sample was measured up to 10 GPa (full spectral analysis shown in Figure \ref{SI:fig:5}. The low-pressure cubic autoclave precursor material was also loaded in a DAC equipped with 400 $\mu$m culets and measured under pressure with argon as a pressure transmitting medium. The Raman spectrum of this sample was measured as a function of pressure to enable comparison between the the laser heated samples.
\\

\paragraph*{Fourier transform infrared spectroscopy:}
Fourier-transform infrared (FTIR) spectra were obtained using a Bruker Vertex spectrometer with Hyperion microscope with MCT detector. The total spectral range for the measurements was $\sim$500 to 8,300 cm$^{-1}$ with a resolution of 4 cm$^{-1}$. For measurements of bulk Mg$_{2}$IrH$_{5}$ with strongly absorbing Ir--H stretching modes, the product powder was mixed with dry KBr ($\sim$5 vol\% hydride) and pressed into a $\sim$1 mm thick pellet.
\\

\paragraph*{DFT calculations:}
Geometry optimizations and all subsequent calculations performed using \textsc{castep} \cite{2005_CASTEP} used the Perdew-Burke-Ernzerhof generalized gradient approximation for solids (PBEsol) \cite{2009_Perdew}, a 600 eV plane-wave cutoff, a $k$-point spacing of 2π× 0.03 \AA$^{-1}$, and default \textsc{castep} norm-conserving  pseudopotentials (NCP) defined with the following strings:\\

\hspace{5.5cm}\parbox{\textwidth}{%
\texttt{
H  1|0.8|14|16|19|10N(qc=8)\\
Mg 1|1.8|3|4|4|30N:31L:32N\\
Ir 3|1.6|8|10|21|50N:60N:51NN:52NN(qc=7)\\
}
}
\\

Structures were converged to give forces and stresses to within 0.05 eV/\AA~and 0.01 GPa.
\\

\paragraph*{Phonon and Raman calculations:}
Gamma-point phonon and Raman intensity calculations were performed using density functional perturbation theory implemented in \textsc{castep}.\cite{2006_Refson}
\\

\paragraph*{VCNEB calculations:}
The phase transition path from the $Amm$2 and $P$4/$nmm$ to $Fm\overline{3}m$ Mg$_{2}$IrH$_{5}$ was evaluated using the The Variable-Cell Nudged Elastic Band (VCNEB) method as implemented in \textsc{uspex} code \cite{VCNEB,USPEX}, using variable elastic constants between 3 and 6 eV/\AA$^{2}$ and a variable number of images between the endpoints. The initial number of images was set to 50, the maximum number of steps 1000, and a convergence threshold on the root mean square of the forces of 10 meV/\AA~was chosen. The energy and forces were calculated using \textsc{vasp} with an energy cutoff of 800 eV on the plane-wave expansion, and for $k$-space integration, a resolution of 0.1 in units of 2$\pi$/\AA~and a Methfessel--Paxton smearing with a width of 0.04 eV. After successful determination of a converged transition path, energies of all structures were recomputed within \textsc{castep} and the most converged setting as described in the previous section.
\\

\clearpage

\subsection*{Thermogravimetric analysis}
\begin{figure}[h!]
    \centering
    \includegraphics[width = 0.8 \linewidth]{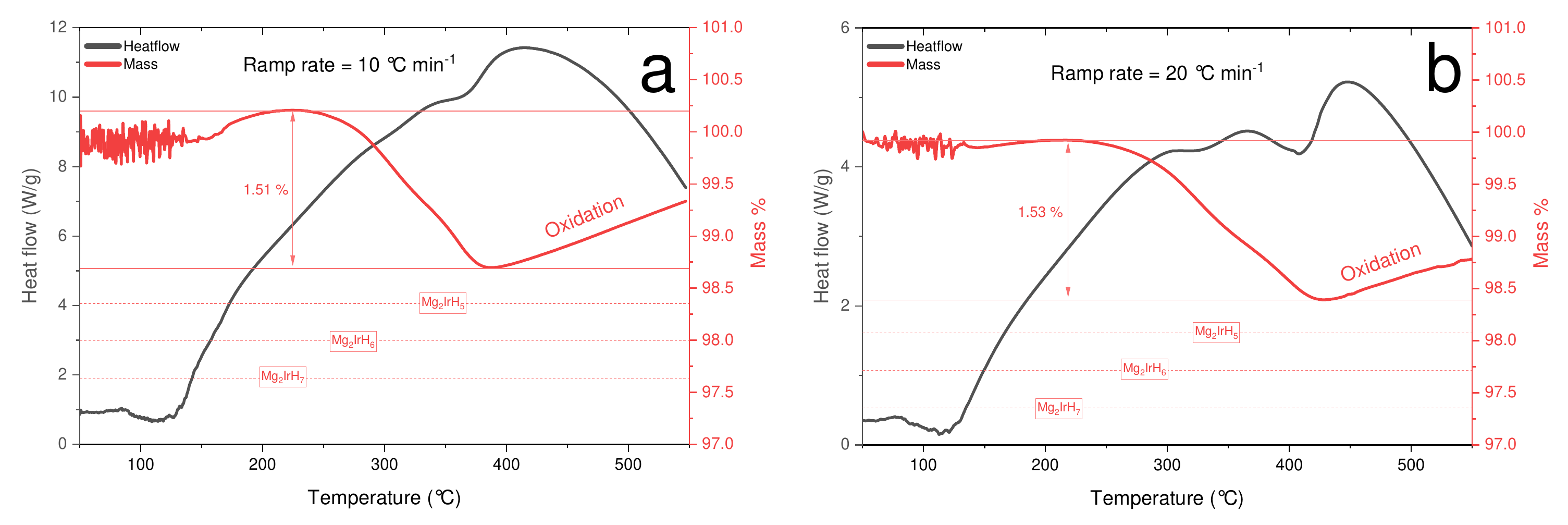}
    \caption{TGA/DSC measurements performed on pellets of Mg$_{2}$IrH$_{5}$ synthesized at $\sim$150 bar of hydrogen in an autoclave system. \textbf{a:} Measurement performed with a ramp rate of 10~\textdegree{}C min$^{-1}$ for a 9.919 mg pellet. \textbf{b:} Measurement performed with a ramp rate of 20~\textdegree{}C min$^{-1}$ for a 16.890 mg pellet. The theoretical mass loss for hydrogen evolution from Mg$_2$IrH$_x$ phases, corrected for 90\% phase purity of the samples, is indicated by the dashed lines.}
    \label{SI:fig:TGA}
\end{figure}

To more directly assess the hydrogen content of bulk samples, we performed thermogravimetric analysis (TGA) on samples synthesized at $\sim$150 bar using the autoclave-type apparatus. The measurements were done under a flow of 500 ml/min Argon (99.99\% purity). In Figure \ref{SI:fig:TGA} we show the TGA data as well as differential scanning calorimetry (DSC) data for two pellets of the same precursor material, run with a ramp of 10~\textdegree{}C min$^{-1}$ and 20~\textdegree{}C min$^{-1}$, respectively. In both measurements the onset of mass loss begins around 250~\textdegree{}C. At higher temperatures the samples begin to gain mass, which is due to unavoidable oxidation, as confirmed by the formation of MgO based on XRD analysis of the recovered product. The graphs show the theoretically expected mass loss for Mg$_{2}$IrH$_{5}$, Mg$_{2}$IrH$_{6}$, and Mg$_{2}$IrH$_{7}$. Sample oxidation during heating precludes definitive compositional analysis, however the TGA measurements remain consistent with Mg$_{2}$IrH$_{5}$.

\subsection*{Semi-quantitative elemental analysis - Energy dispersive x-ray spectroscopy}

\begin{figure*}[h!]
    \centering
    \includegraphics[width = 0.8 \linewidth]{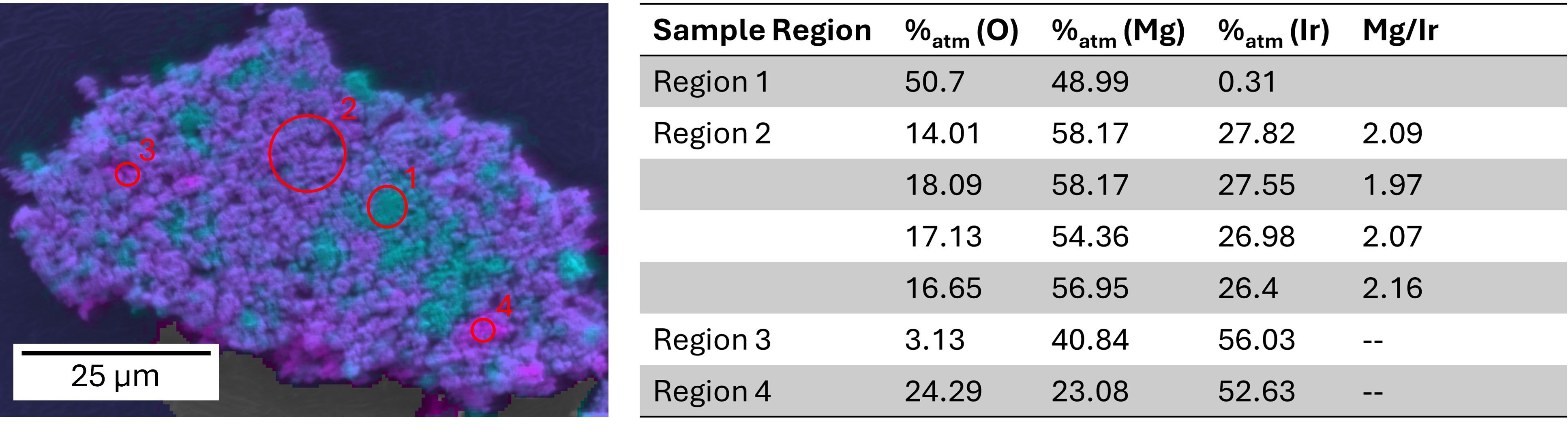}
    \caption{SEM micrograph color coded with EDXS intensities from Mg (purple), Ir (pink), and O (teal). Four regions are highlighted with red circles and the semi-quantitative atomic percentages obtained from spectra measured in the respective regions are summarized in the table. Region 1 represents MgO impurities while other regions are consistent with a composition of 2Mg:1Ir}
    \label{SI:fig:EDX}
\end{figure*}

To independently probe the composition of the Mg$_{2}$Ir metal sub-lattice in the synthesized cubic compound, we performed energy-dispersive X-ray spectroscopy (EDXS) measurement to provide semi-quantitative elemental analysis. A powder sample synthesized in an autoclave (nominally Mg$_{2}$IrH$_{5}$) was compressed between two 1000 $\mu$m diamonds to form a pellet with a flat surface. A piece of this pellet was transferred to an aluminium stub with carbon tape and was measured in a Zeiss Auriga field emission SEM operated at 15 kV and a beam current of $\sim$1 nA. Data were collected using an Oxford X-Max EDXS system running the Aztec software. We measured EDXS maps to establish the presence of minority phases, which are spatially distinguished in Figure \ref{SI:fig:EDX}. The Mg:Ir atomic ratios measured in different regions are shown in Figure \ref{SI:fig:EDX}.

\subsection*{Transport measurements}
The temperature-dependent resistance of a $\sim$ 90\% phase pure pellet of  Mg$_{2}$IrH$_{5}$ was measured between 4--300 K using a linear four-probe geometry using a Physical Properties Measurement System. The sample contains $\sim$10\% unreacted metallic Ir and MgIrH$_x$ ($Pm\overline{3}m$) impurities, giving rise to complex behaviour, however, the overall temperature dependence is consistent with the presence of a band gap in the Mg$_{2}$IrH$_{5}$ dominant phase.

\begin{figure*}[h!]
    \centering
    \includegraphics[width = 0.8 \linewidth]{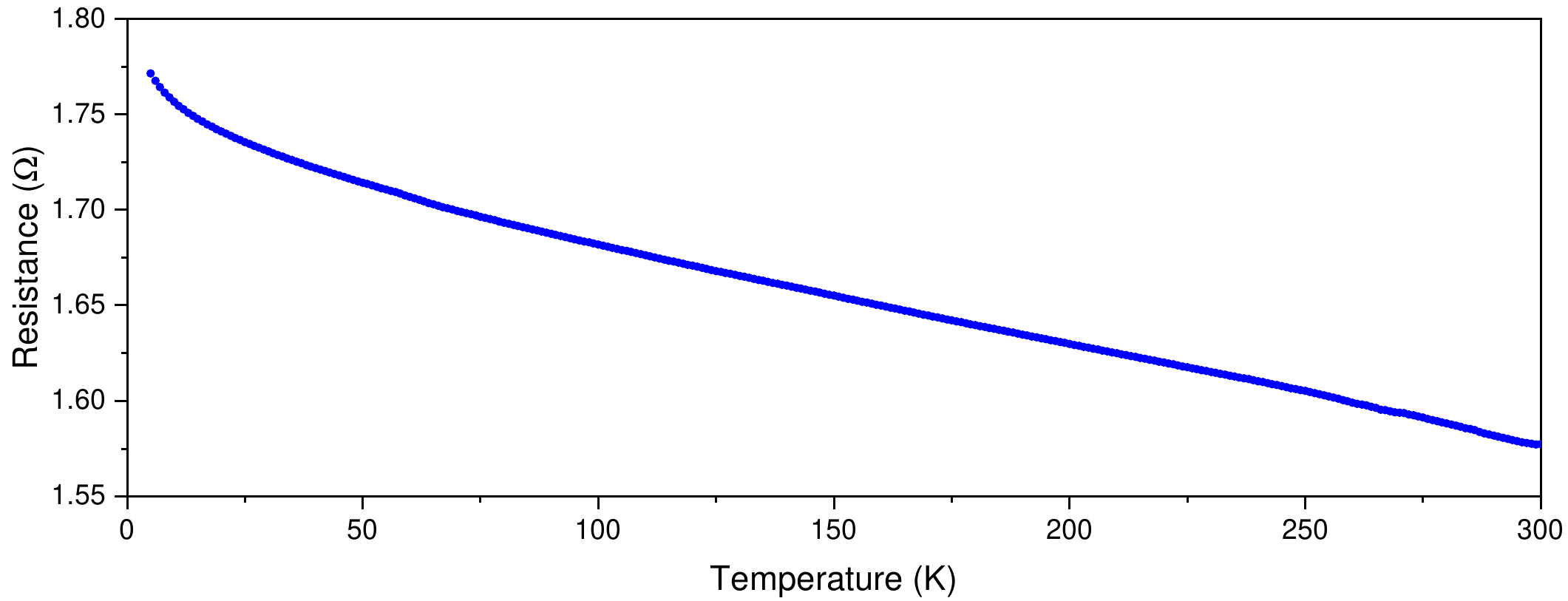}
    \caption{Four-probe resistance measurement of a pellet of Mg$_{2}$IrH$_{5}$.}
    \label{SI:fig:resistivity}
\end{figure*}

\subsection*{Model structures for Mg$_{2}$IrH$_{5}$}
\begin{figure}[h!]
    \centering
    \includegraphics[width = 0.7 \linewidth]{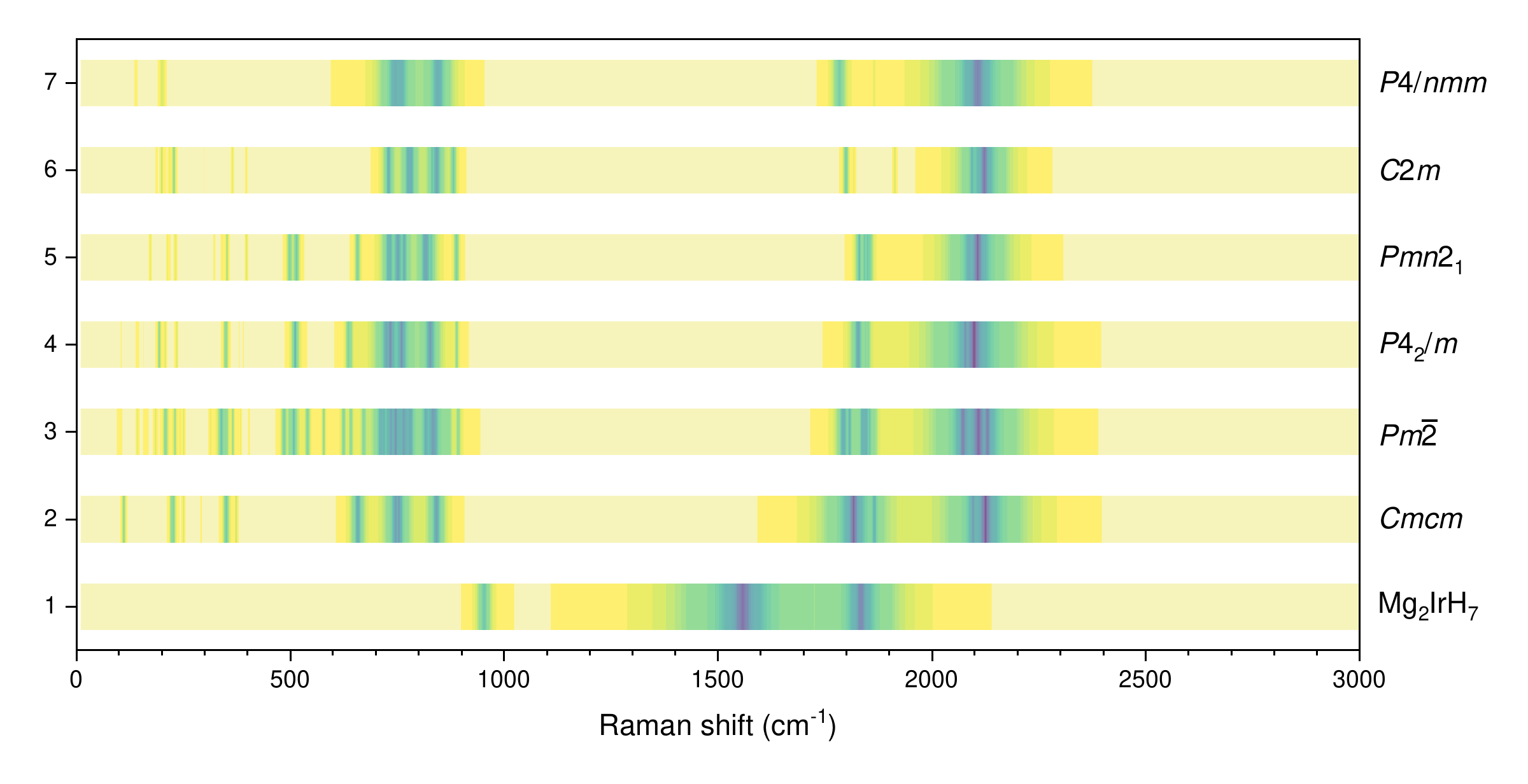}
    \caption{Calculated Raman modes of different ordered approximations of the cubic Mg$_{2}$IrH$_{5}$ compound as well as Mg$_{2}$IrH$_{7}$. Intensities (color scale) are plotted on a log scale to better visualize all modes.}
    \label{SI:fig:2}
\end{figure}

As mentioned in the main text, ordered structures were used as approximations for phonon calculations of Mg$_{2}$IrH$_{5}$, which is disordered in space group $Fm\overline{3}m$. Six of the lowest-energy static approximations of Mg$_2$IrH$_5$ established by Dolui et al. \cite{2023_Kapildeb} were used as models for Raman and Infrared calculations using \textsc{CASTEP}. These structures are provided as .cif files and their calculated Raman mode frequencies and intensities are shown in Figure \ref{SI:fig:2}. 

\begin{figure*}[h!]
    \centering
    \includegraphics[width = 0.65 \linewidth]{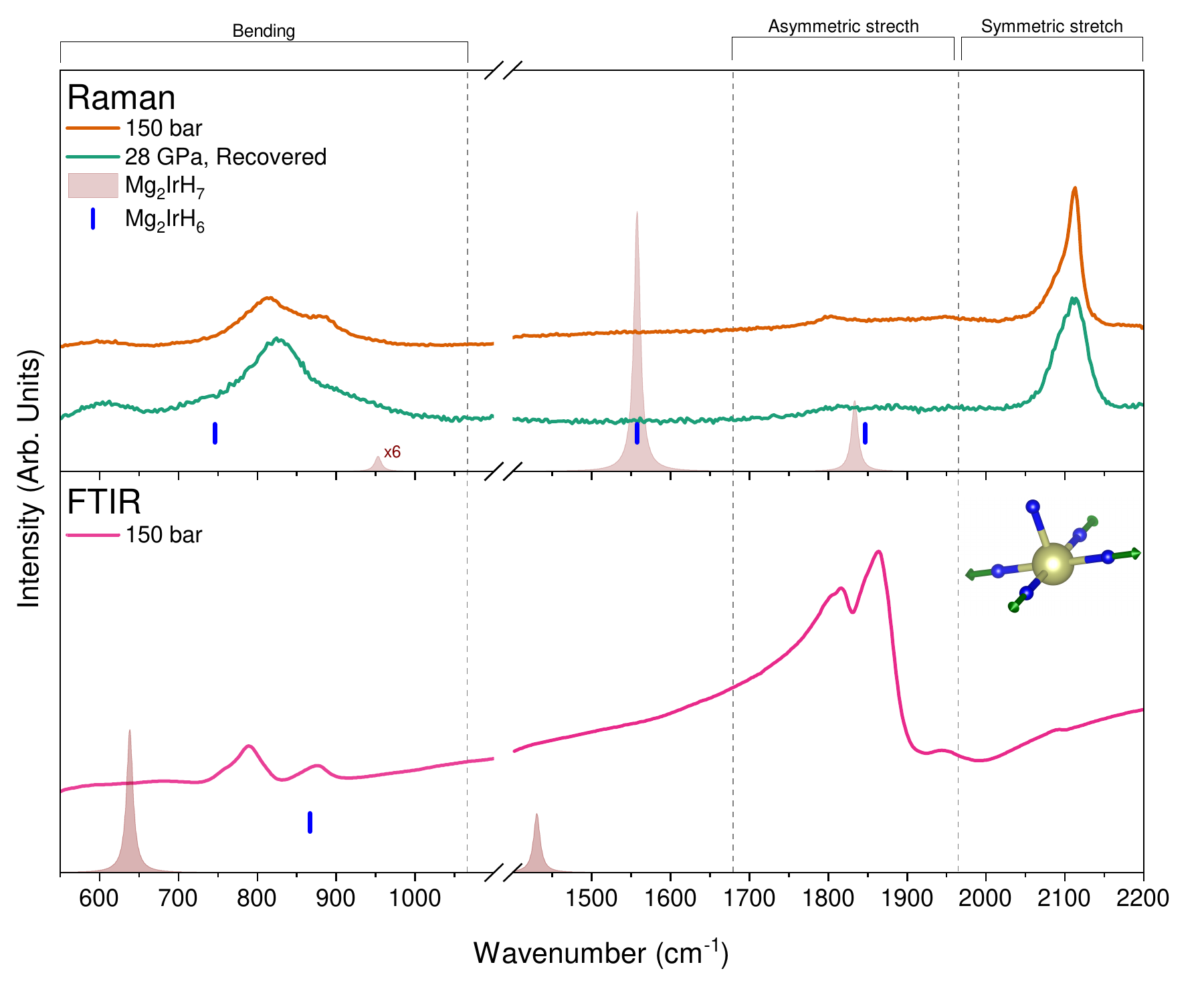}
    \caption{Spectroscopic data for FCC samples produced at different $P$--$T$ conditions. %Mg$_{2}$IrH$_{5}$. 
    Raman spectra of FCC samples taken at ambient pressure for samples synthesized at 150 bar and 28 GPa, as well as the FTIR spectrum of a sample synthesized at 150 bar. All spectra shown are at ambient pressure. The region from 1200--1500 cm$^{-1}$ is removed due to Raman signal from the diamonds and has no other observed peaks. As clearly seen here, the compounds Mg$_{2}$IrH$_{6}$ and Mg$_{2}$IrH$_{7}$ are both poor matches with the experimental data. The axes limits on the figure are chosen for direct comparison with Figure 3 in the main text.}
    \label{SI:fig:3}
\end{figure*}

\begin{figure*}[ht!]
    \centering
    \includegraphics[width = 0.65 \linewidth]{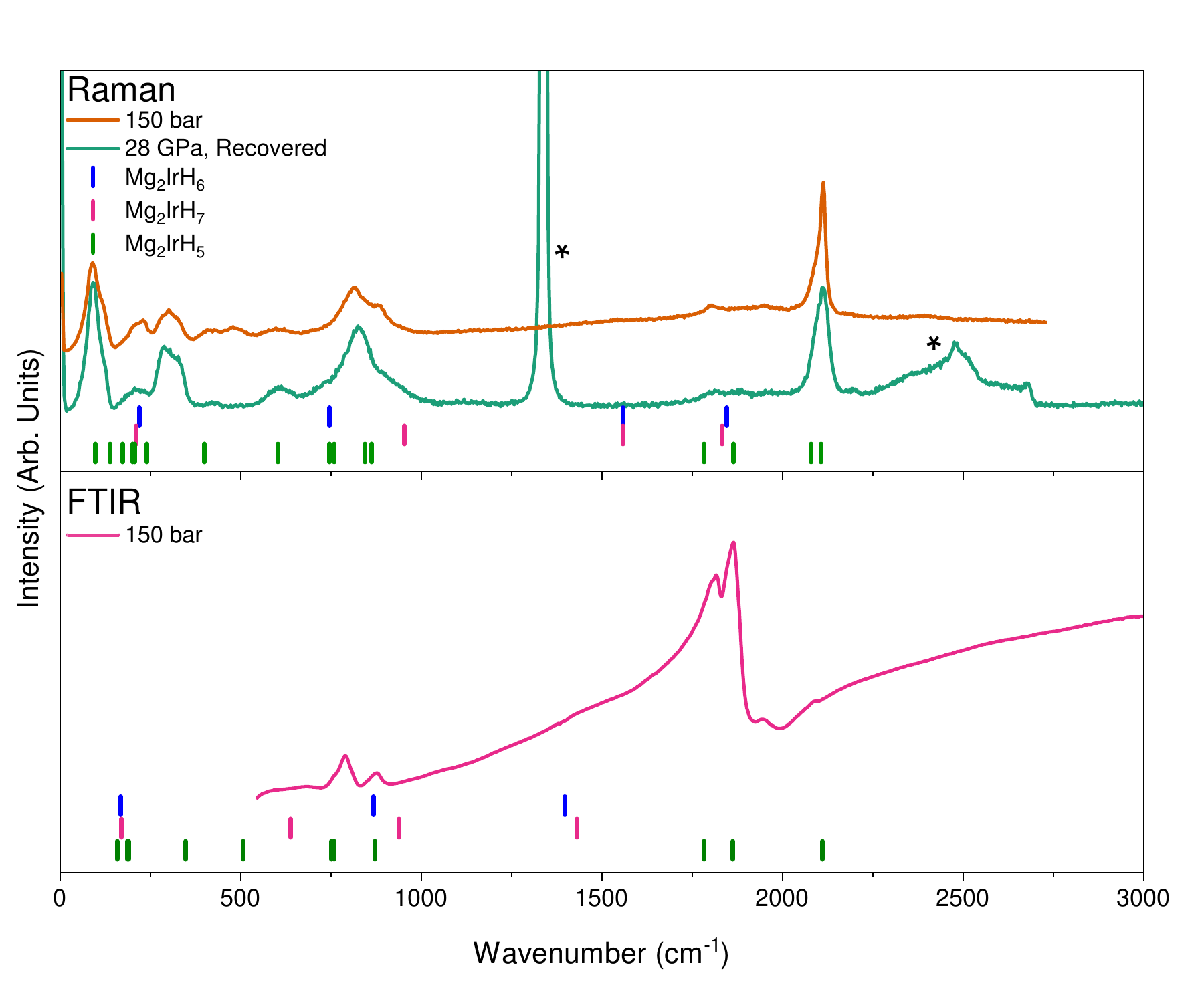}
    \caption{Full-range spectroscopic data for FCC samples produced at different $P$--$T$ conditions. Stars indicate peaks coming from diamond.}
    \label{SI:fig:4}
\end{figure*}

For all of the ordered Mg$_2$IrH$_5$ strutural models, the molecular vibrations, such as the symmetric Ir--H stretching modes near $\sim$2100 cm$^{-1}$ and the Ir--H bending modes at $\sim$800 cm$^{-1}$ are consistent across all of the approximations, showing that these ordered models capture the molecular vibrations well, irrespective of the specific crystal symmetry. The lower-energy phonon modes below $\sim$500 cm$^{-1}$ show a stronger dependence on the specific crystal structure. For comparison, Mg$_{2}$IrH$_{7}$ is also plotted and is significantly different from all Mg$_{2}$IrH$_{5}$ models. Figure \ref{SI:fig:3} shows the experimental spectra from Figure 3 in the main text, but with the calculated Mg$_{2}$IrH$_{7}$ frequencies and intensities overlaid (Mg$_{2}$IrH$_{6}$ is similar but intensities are not calculated due to the metallic nature). Full-range spectra (including contributions from diamond) are shown in Figure \ref{SI:fig:4} with tick marks representing the calculated frequencies for Mg$_{2}$IrH$_{5}$, Mg$_{2}$IrH$_{6}$, and Mg$_{2}$IrH$_{7}$.

\begin{figure}[h!]
    \centering
    \includegraphics[width = 0.65 \linewidth]{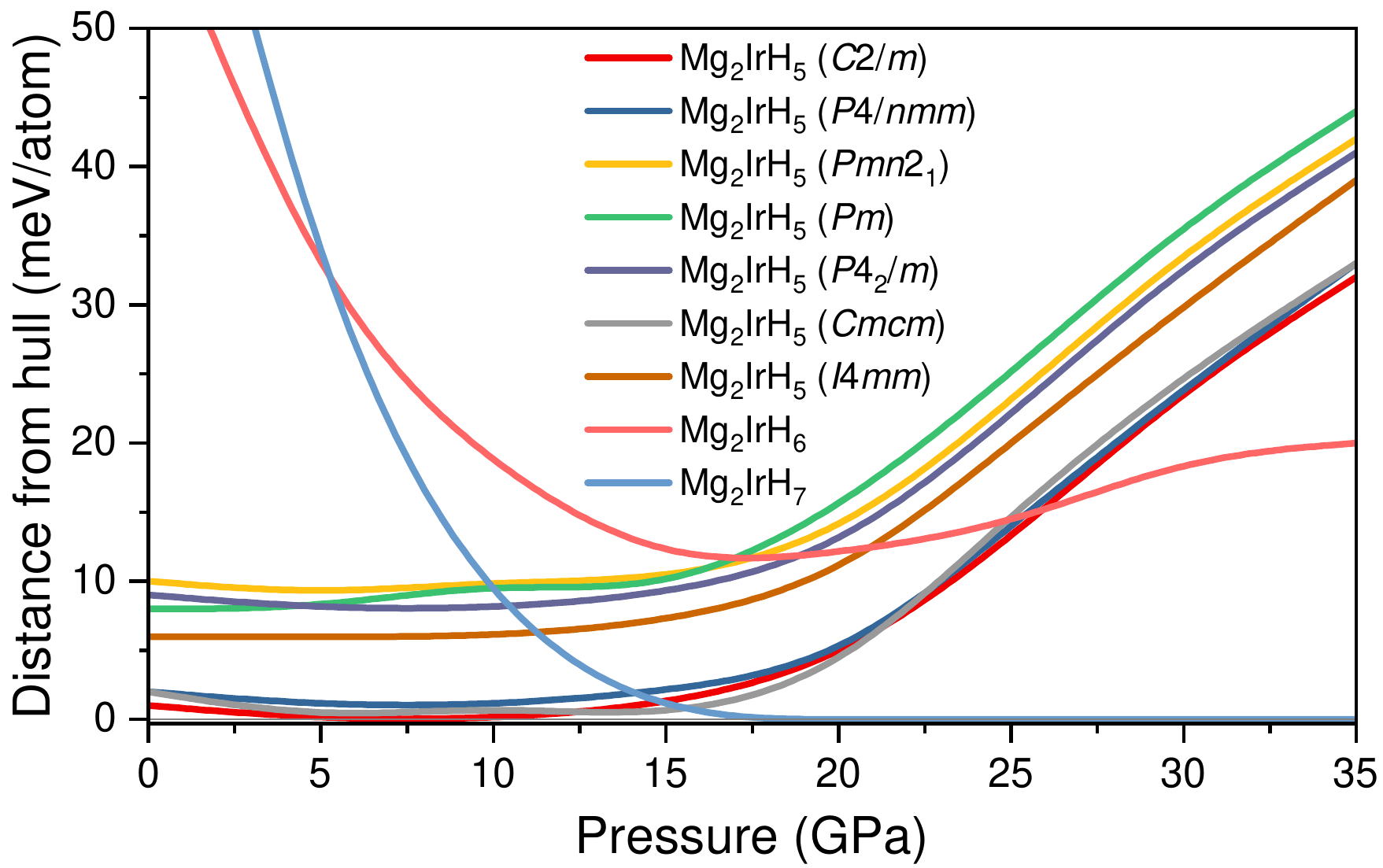}
    \caption{Pressure dependence of the convex hull distance for several lower-symmetry approximations of Mg$_{2}$IrH$_{5}$ as well as FCC Mg$_{2}$IrH$_{6}$ and Mg$_{2}$IrH$_{7}$}
    \label{SI:fig:hull}
\end{figure}

All the structural approximations of Mg$_{2}$IrH$_{5}$ used in this study are available in the SM as *.cif files. To examine the energies of these phases, we calculated the hull distances for all Mg$_{2}$IrH$_{5}$ approximations as a function of pressure, shown in Figure \ref{SI:fig:hull}. All Mg$_{2}$IrH$_{5}$ approximations show qualitatively similar energies with similar pressure dependence. At 0 GPa, $P$4/$nmm$ Mg$_2$IrH$_5$ is calculated to be 2 meV/atom above the hull. 

\begin{figure}[h!]
    \centering
    \includegraphics[width = 0.65 \linewidth]{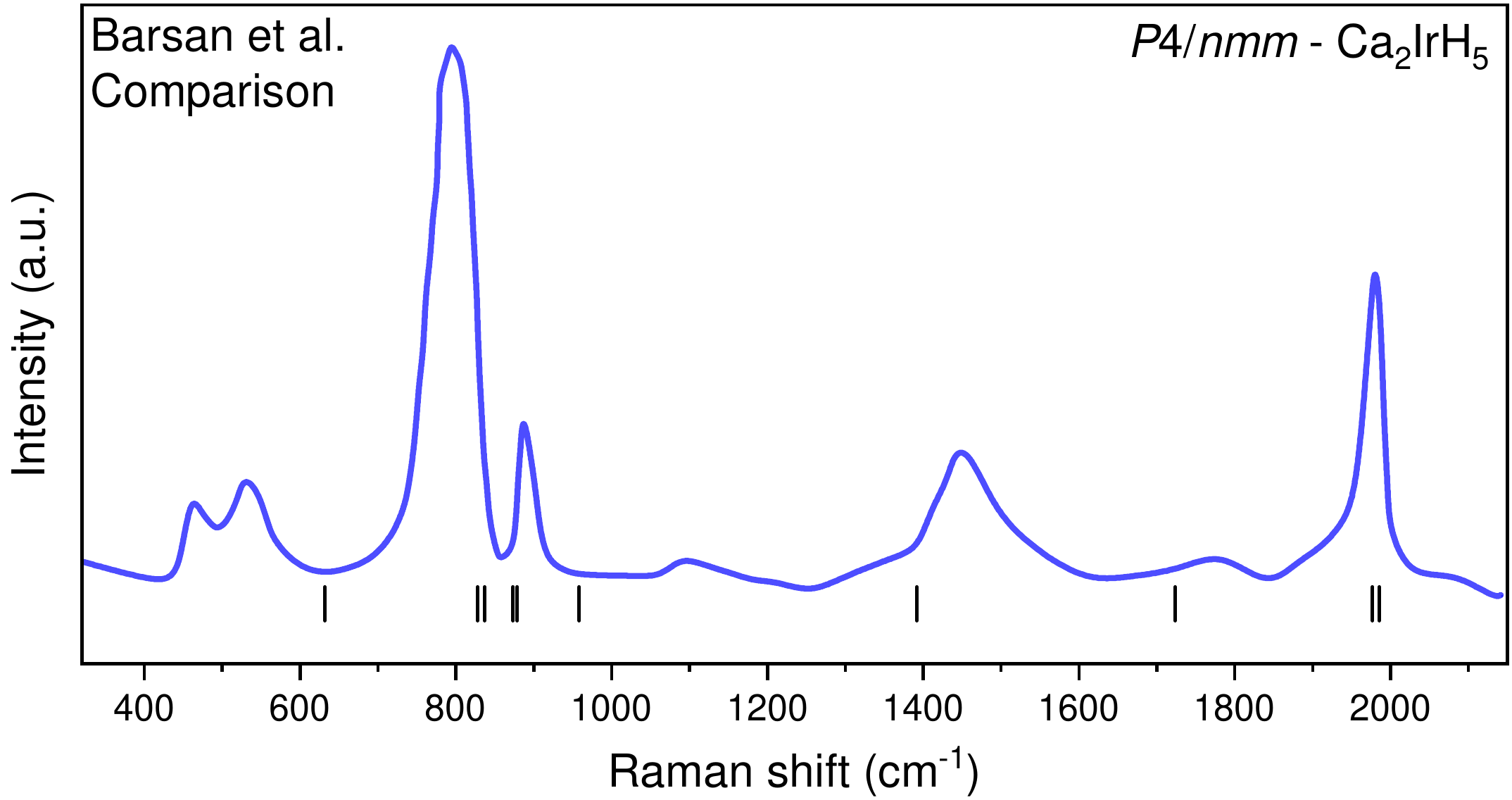}
    \caption{Calculated Raman modes for Ca$_{2}$IrH$_{5}$ compared with digitized spectrum from Barsan et al. 
\cite{Barsan2012} Tick marks are the frequencies obtained from our calculations.}
\label{SI:fig:barsan}
\end{figure}

To benchmark our approximations we calculated the energies of the Raman modes for Ca$_{2}$IrH$_{5}$, which was previously reported by Barsan et al. In Figure \ref{SI:fig:barsan} we have plotted the experimental data obtained by Barsal et al. \cite{Barsan2012} (digitized for comparison) and our calculations. We see that the symmetric stretching modes of of the IrH$_{5}$ units in Ca$_{2}$IrH$_{5}$ are well reproduced along with the asymmetric stretching modes. The low-energy modes are not as well reproduced, which is what we expect from our approach due to differences in the lattice symmetry.

\clearpage

\begin{figure}[h!]
    \centering
    \includegraphics[width = 0.65 \linewidth]{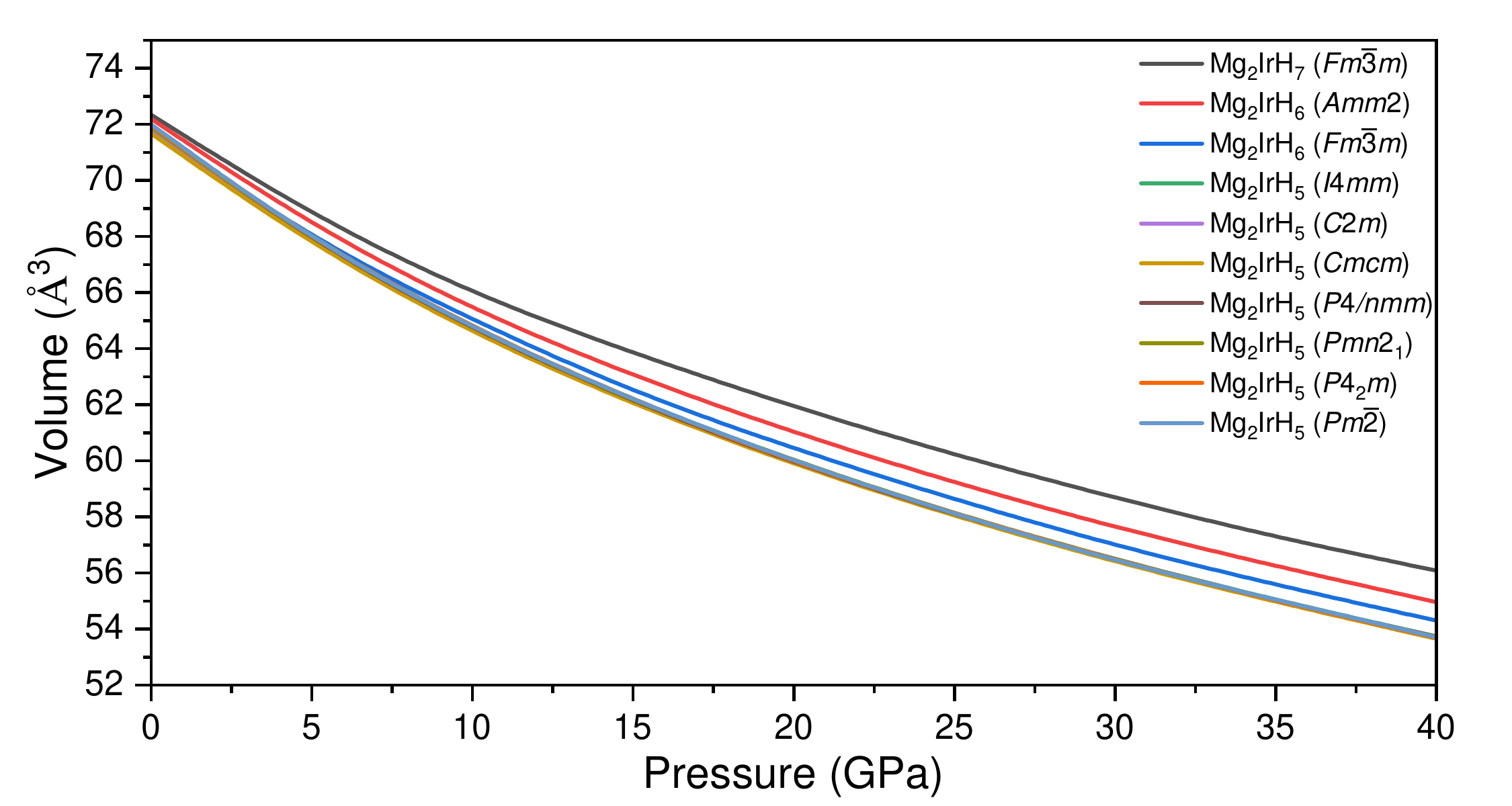}
    \caption{The calculated unit cell volume of Mg$_{2}$IrH$_{7}$, Mg$_{2}$IrH$_{6}$, and several ordered approximations of disordered FCC Mg$_{2}$IrH$_{5}$ as a function of pressure. The volumes of all Mg$_{2}$IrH$_{5}$ structures are essentially on the same line.}
\label{SI:fig:EOS_calc}
\end{figure}

Furthermore, the unit cell volumes as a function of pressure were also calculated for the ordered Mg$_{2}$IrH$_{5}$ approximations, which is shown in Figure \ref{SI:fig:EOS_calc}. The calculated volumes for these structrures are essentially indistinguishable in the plot, but show differences from FCC Mg$_{2}$IrH$_{6}$, FCC Mg$_{2}$IrH$_{7}$ and $Amm$2 FCC Mg$_{2}$IrH$_{6}$.

\subsection*{Pressure dependence of the low energy Raman modes}
\begin{figure}[h!]
    \centering
    \includegraphics[width = 0.45 \linewidth]{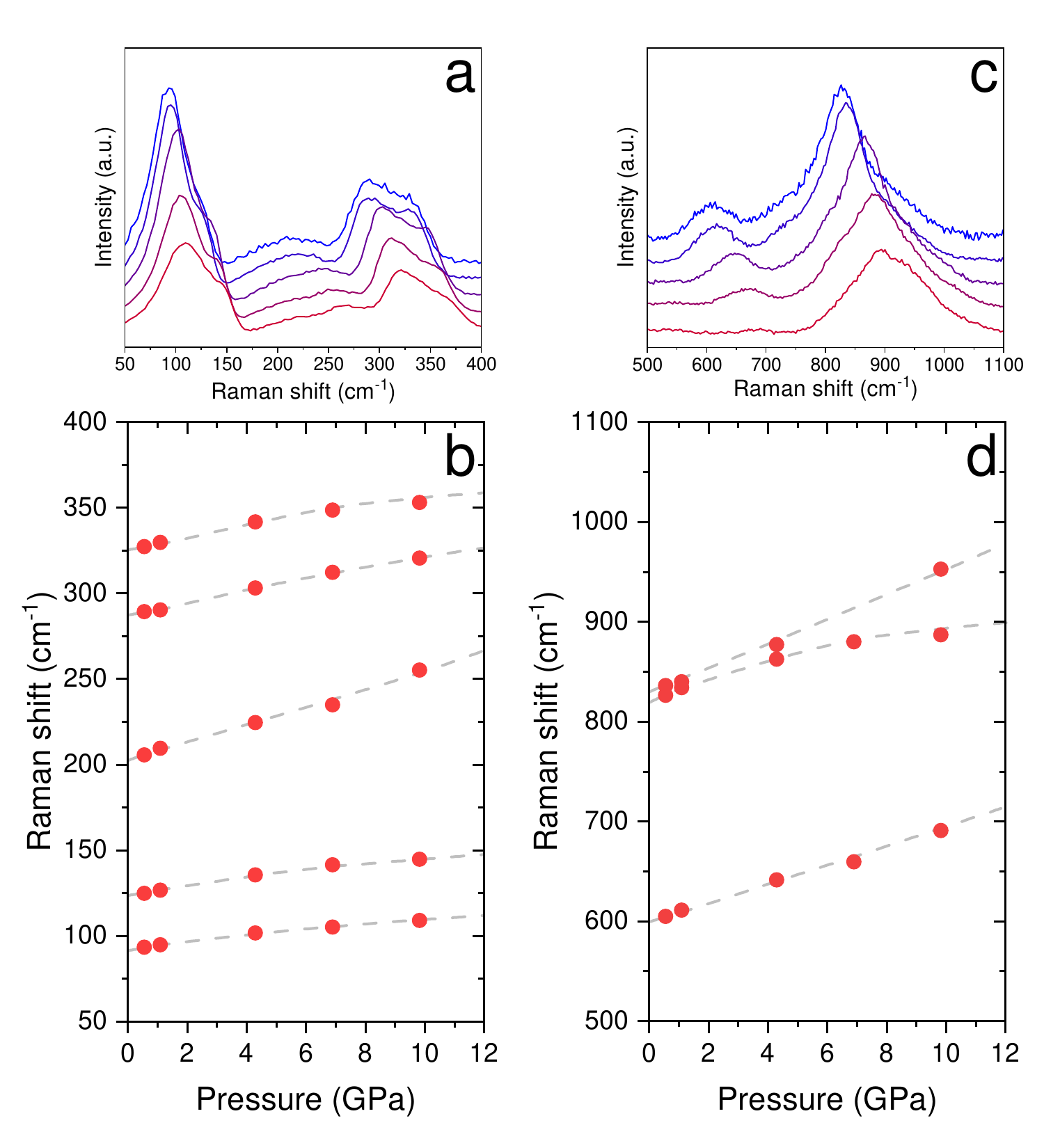}
    \caption{The pressure dependence of the low-frequency Raman peaks up to 10 GPa as measured for a sample synthesized at 28 GPa in a DAC. \textbf{a:} Measured spectra in the range 50--400 cm$^{-1}$. \textbf{b:} Fitted modes in the range 50--400 cm$^{-1}$. \textbf{c:} Measured spectra in the range 500--1100 cm$^{-1}$. \textbf{d:} Fitted modes in the range 500--1100 cm$^{-1}$}.
    \label{SI:fig:5}
\end{figure}

We measured the Raman modes corresponding to the lower energy lattice vibrations as well as the bending modes of the Ir--H units. These were measured up to 10 GPa on a sample which was recovered from laser heating at 28 GPa in a DAC with hydrogen as a pressure transmitting medium and reloaded in a DAC with Argon as a pressure transmitting medium to eliminate scattering from molecular hydrogen. The modes were fitted with the apparent number of Lorentzian peaks. Figure \ref{SI:fig:5}a shows the spectral range between 50--400 cm$^{-1}$ and \ref{SI:fig:5}b shows the peak positions obtained from fitting peaks in the same region. Figure \ref{SI:fig:5}c shows the spectral range between 500--1100 cm$^{-1}$ and \ref{SI:fig:5}d shows the peak positions obtained from fitting peaks in the same region.

\subsection*{Compressibility of $Pm\overline{3}m$ phase}
\begin{figure}[ht!]
    \centering
    \includegraphics[width = 0.65 \linewidth]{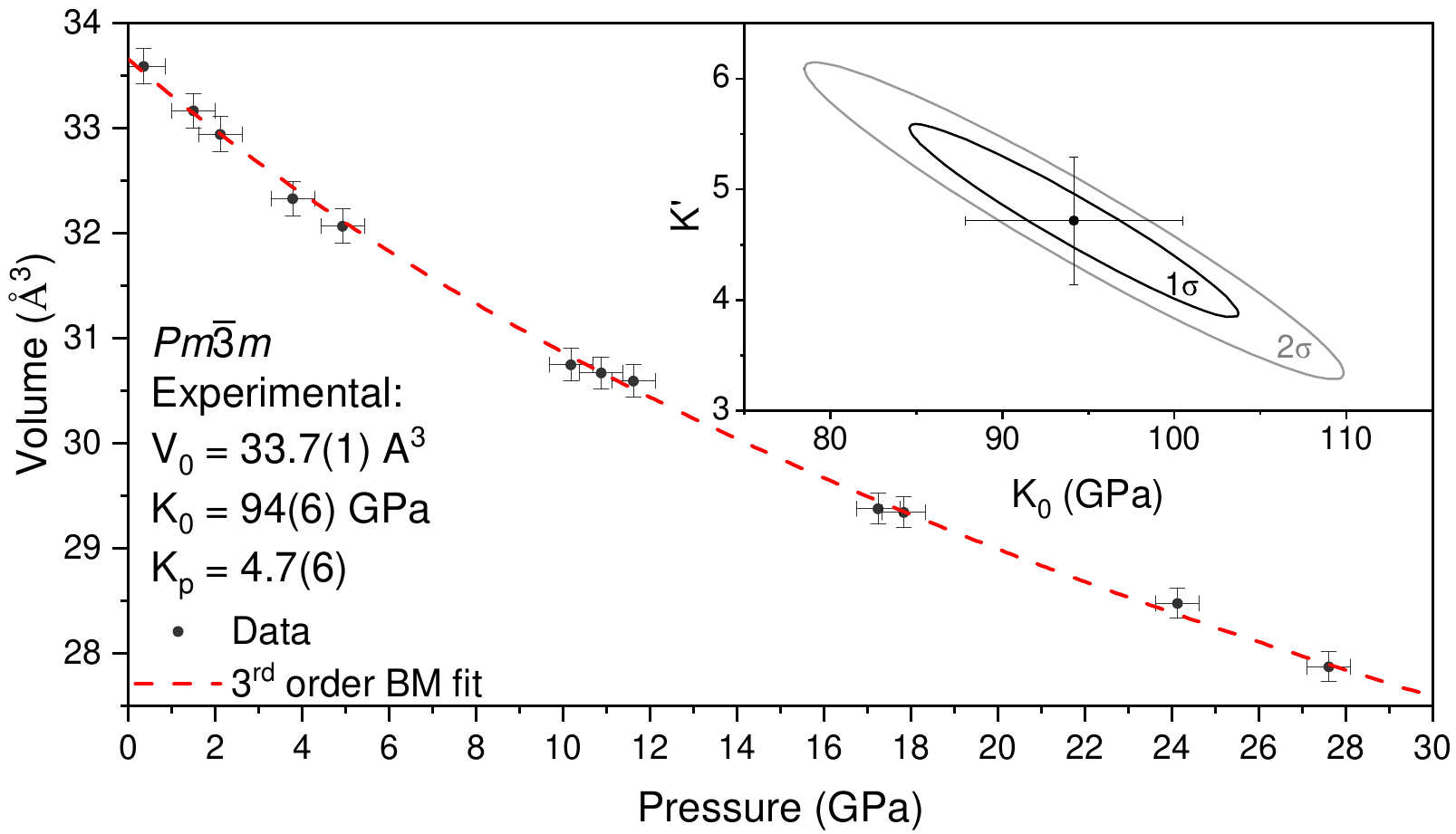}
    \caption{Unit cell volume of $Pm\overline{3}m$ MgIrH$_x$ minority phase present in samples before heating. The data is fitted using a third-order Birch Murnaghan EOS and the obtained fitting parameters are given. The inset shows the correlation ellipses for K$_{0}$ and K' at one and two standard deviations.}
    \label{SI:fig:alloy}
\end{figure}

As a minor phase in the hydride precursor material produced using the autoclave system, we observed a cubic phase with $Pm\overline{3}m$ symmetry. A similar impurity assigned as an `MgIr' alloy was reported in the thesis of Bonhomme. \cite{1995_Bonhomme_thesis} Here, we assign the composition as MgIrH$_x$, where $x$ is unknown and possibly zero, based on DFT volume predictions that range between MgIr and MgIrH$_3$. While unable to definitively solve the structure, we followed the unit cell parameters as a function of pressure upon decompression to obtain the equation of state, as seen in Figure \ref{SI:fig:alloy}. We fitted the volume as a function of pressure using a third-order Birch Murnaghan EOS and in the inset of the figure we provide the correlation ellipses for K$_{0}$ and K'.

\begin{figure}[ht!]
    \centering
    \includegraphics[width = 0.65 \linewidth]{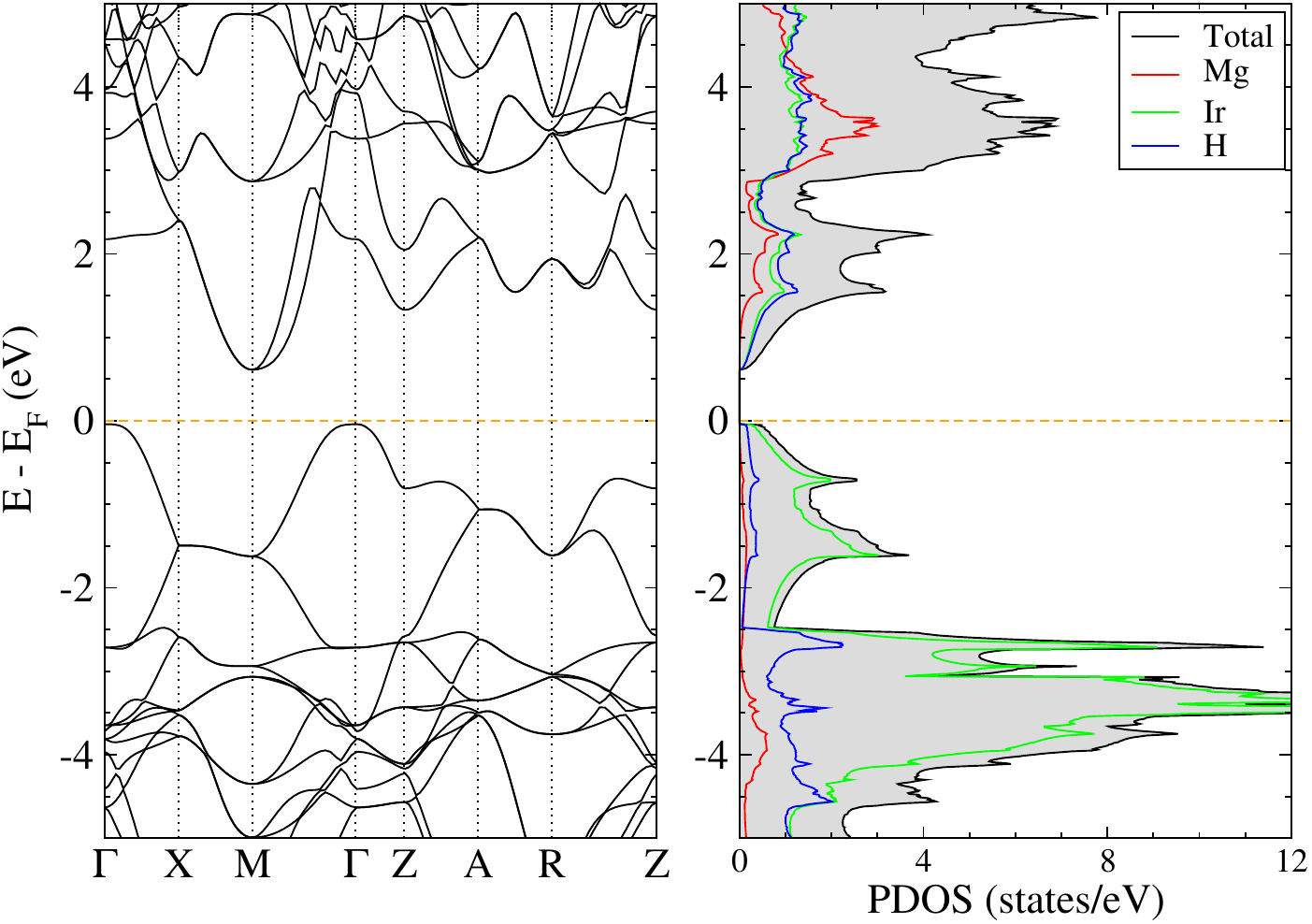}
    \caption{In the left panel, DFT-GGA calculated electronic band structure of $P4/nmm$ Mg$_2$IrH$_5$~at ambient pressure. In the right panel, the total electron density of states (DOSs) projected into Mg (red), Ir (green), and H (blue) atoms.}
    \label{SI:fig:H5-bands:}
\end{figure}

\clearpage

%\newpage
%\bibliography{bib}
%merlin.mbs apsrev4-1.bst 2010-07-25 4.21a (PWD, AO, DPC) hacked
%Control: key (0)
%Control: author (8) initials jnrlst
%Control: editor formatted (1) identically to author
%Control: production of article title (-1) disabled
%Control: page (0) single
%Control: year (1) truncated
%Control: production of eprint (0) enabled

%

\end{document}